\documentstyle[11pt,psfig,aasms]{article}
\tightenlines
\def\cm{\,{\rm cm}}
\def\kms{\mbox{km\,s$^{-1}$}}
\def\Msol{\mbox{M$_\odot$}}
\def\bigexpec#1{\biggl \langle#1\biggl \rangle}
\def\expec#1{\langle#1\rangle}
\def\dd{\,{\rm d}}
\newcommand{\op}{Ly$\alpha$\ }
\newcommand{\hi}{\mbox{H{\scriptsize I}}}
\newcommand{\heii}{\mbox{He{\scriptsize II}}}
\newcommand{\cii}{\mbox{C{\scriptsize II}}}
\newcommand{\ciii}{\mbox{C{\scriptsize III}}}
\newcommand{\civ}{\mbox{C{\scriptsize IV}}}
\newcommand{\cv}{\mbox{C{\scriptsize V}}}
\newcommand{\siiv}{\mbox{Si{\scriptsize IV}}}
\newcommand{\nv}{\mbox{N{\scriptsize V}}}
\newcommand{\ovi}{\mbox{O{\scriptsize VI}}}
\newcommand{\mgii}{\mbox{Mg{\scriptsize II}}}

\def\aua#1#2{{ #1, }{\aap,}{#2}}
\def\mapj#1#2{{#1, }{\apj,} {#2}}
\def\mapjs#1#2{{#1, }{\apjs,} {#2}}
\def\maj#1#2{{#1, }{\aj,} {#2}}
\def\mmnras#1#2{{#1, }{\mnras,} {#2}}
\def\nat#1#2{{#1, }{Nat,} {#2}}

\begin{document}

\begin{titlepage}   

\today

\title{QSO Metal Absorption Systems at High Redshift 
and the Signature of Hierarchical Galaxy Formation}

\author{Michael Rauch\altaffilmark{1,4}, Martin G. Haehnelt\altaffilmark{2},and
Matthias Steinmetz\altaffilmark{2,3}
}
\altaffiltext{1}{Astronomy Department, California Institute of Technology,
Pasadena, CA 91125, USA}
\altaffiltext{2}{Max-Planck-Institut f\"ur Astrophysik, Postfach 1523, 85740
Garching, Germany}
\altaffiltext{3}{Department of Astronomy, University of California,  Berkeley,
CA 94720, USA}
\altaffiltext{4}{Hubble Fellow}

\centerline{mr@astro.caltech.edu, mhaehnelt@mpa-garching.mpg.de,
msteinmetz@astro.berkeley.edu}
\vskip 1.5cm

{Subject Headings:  cosmology: theory, observation --- galaxies:
formation, evolution --- intergalactic medium --- quasars: absorption lines}

\vskip 1.5cm

\abstract{
In a hierarchical cosmogony  galaxies build up by
continuous merging of  smaller structures. At redshift three the matter
content of a typical present-day galaxy is dispersed over several
individual clumps embedded in sheet-like structures,
often aligned along filaments. We have used hydrodynamical simulations
to investigate the spatial distribution and absorption properties of
metal enriched gas in such regions of ongoing galaxy formation.  The
metal and hydrogen absorption features produced by the collapsing
structures closely resemble observed QSO absorption systems over a wide
range in \hi\ column density. Strong \cii\ and \siiv\ absorption occurs for
lines-of-sight passing the densest regions close to the center of the
protogalactic clumps, while \civ\  is a good tracer of the prominent
filamentary structures and \ovi\ becomes the strongest absorption feature
for lines-of-sight passing through low-density regions far away from
fully-collapsed objects. The observed column density ratios of the
different ionic species  can be well reproduced if a mean metallicity
[Z/H] =$-2.5$,  relative abundances as found in metal-poor stars,
a UV background with intensity $J_{-22}=3$ at the Lyman limit, and
either a power law spectrum ($J\propto\nu^{-1.5}$) or the
spectral shape proposed by Haardt \& Madau (1996)  are
assumed.  The observed scatter in [C/H] is about a magnitude larger
than that in the  simulations suggesting an inhomogeneous metal
distribution. Observed and simulated  Doppler parameter distributions
of  \hi\ and \civ\ absorption lines are in good agreement 
indicating that shock heating due to gravitational collapse is a 
second important heating agent in addition to photoionization
heating.  The two-point correlation function
of \civ\ lines agrees reasonably well with the observed correlation
function suggesting that its high-velocity tail is caused by
occasional alignments of several absorbing clumps still expanding 
with the Hubble flow.  Both high-ionization multi-component 
heavy-element absorbers and
damped Lyman alpha systems  can arise from groups of moderately sized
protogalactic clumps ($M_{\rm baryon}$$\sim 10^9 M_{\odot}$). Recent 
detections of star forming galaxies at similar redshifts are consistent
with this picture.}

\end{titlepage}

\section{Introduction}

Can we understand galaxy formation at high redshift studying QSO
absorption systems?  Lines-of-sight (LOS) to distant QSOs must frequently
traverse protogalactic environments. If the gas in these regions
has already experienced some metal enrichment the 
galaxy formation process should produce a characteristic pattern of  metal
absorption lines in the QSO spectrum. If we succeed in identifying
the spectroscopic signature of such regions, high-resolution spectra of
QSOs at high redshift should be able to yield detailed insights into
the physical state of the gas in young galaxies and the gaseous
reservoir from which they form. Once the observable absorption features
have been related to the properties of the absorbing  objects one may
hope to be able to actually constrain the galaxy formation
process and cosmological parameters.  This paper is mostly concerned with
the first aspect:  what {\it are} absorption systems 
(in particular: heavy element systems) at high redshift, and can we
model them realistically by adopting a specific
scenario of structure formation?

QSO absorption studies provide only one-dimensional information, so
observations of single LOS cannot give independently
the spatial density and the size of the objects causing the absorption.
Only the product of number density and cross-section is constrained.
Additional information about the spatial distribution of the absorbing
gas has to be sought, e.g.  by the identification of heavy element
absorbers with a class of galaxies whose number density is known.
As Bahcall \& Spitzer (1969) and Burbidge et al.~(1977) pointed out,
the  cross section of the absorbing gas would have to be  
much larger than typical half-light radii of present-day spiral 
galaxies if these objects had the same comoving space density
as such galaxies. First observational evidence in favor of  large
metal absorption cross-sections came  from work by Bergeron (1986) 
who discovered low redshift galaxies close to the LOS to QSOs with 
known \mgii\ absorption systems, coincident in redshift with the
absorbers.  Surveys of optically thick (Lyman Limit) absorption
systems and galaxies at intermediate redshift have reported results 
consistent with galaxies being surrounded by \mgii\ absorbing halos of 
radius $\sim 40$ kpc (e.g.~Bergeron (1995),  Steidel (1995), 
Churchill, Steidel \& Vogt (1996)).
Extending such work to optically thin systems and lower redshifts 
Lanzetta et al.~(1995) found evidence for even larger \hi\ halos with 
radii of order 160 kpc.  Similarly,  damped  \op systems were
interpreted by Wolfe and collaborators (e.g. Wolfe 1988, 
Lanzetta et al.~1991) as large protodisks, progenitors of present-day 
spiral galaxies with significantly larger radii at high redshift.

The large-halo/disk scenario can qualitatively explain  the component
and internal structure of heavy element absorption systems,
especially the strong clustering measured for \civ\ systems
(Sargent et al.~1979, Sargent, Boksenberg \& Steidel 1988, Petitjean
\& Bergeron 1994). In this
picture the individual absorption components could be clouds orbiting
in a halo or co-rotating in a disk, produced and replenished e.g.  by
thermal instability (Bahcall 1975, Fall \& Rees 1985, Mo 1994, 
Mo \& Miralda-Escud\'e 1996).

However, unambiguous evidence in favor of  large absorption
cross-sections  and the identification of the absorbing  objects with
massive galaxies is restricted to low redshift observations.  Little is
known about the sizes of damped \op absorbers at high redshift (e.g.
M\o ller \& Warren 1995, Djorgovski et al.~1996) and massive disks are
not the only viable dynamical model for the observed velocity
structure. Moreover,  observations of galaxies at high redshift are
strongly biased towards  objects with high current star formation
rates. Thus it is not clear whether the transverse separations
on the sky of galaxy-absorber pairs coincident in redshift do reliably
indicate the presence and sizes of any (hypothetical) halos/disks.
It is difficult to ascertain that there is not an undetected 
(not necessarily faint) galaxy closer to the LOS.

The possibility of ``smaller'' but more numerous objects as sources of
the metal absorption has  been explored earlier.  Tyson (1988) 
suggested identifying damped \op systems with gas-rich dwarf galaxies
instead of large proto-disks.  York et al.~(1986)  discussed
clusters of such objects to explain the component structure of \civ\
systems.  Individual galaxy halos cannot produce potential wells deep
enough to explain the largest velocity splittings of \civ\ 
systems (up to 1000\,\kms) as
virialized motions.  Pettini et al.~(1983) and Morris et
al.~(1986) concluded that the velocity splitting and large
cross-sections are equally difficult to understand if 
objects similar to  present-day
galaxy clusters placed at high redshift were causing the
absorption.

Earlier attempts at understanding QSO absorption systems have mostly
employed heuristic models.  The ionization state of the gas was
calculated  with simplified assumptions about the geometry, temperature
and dynamical structure  of the gas.  Recently, realistic
hydrodynamical simulations of the fate of gas in a universe subject to
hierarchical structure formation (Cen et al.~1994, Hernquist
et al.~1996) have led to a deeper understanding of the large scale 
distribution of baryons. In the new picture a coherent filamentary 
large scale structure of dark matter and baryons is  responsible for 
\op forest absorption lines (e.g. Petitjean, M\"ucket \& Kates 1995; 
Zhang, Anninos \& Norman 1995; Miralda-Escud\'e et al.~1996). 
Denser condensations embedded in the filaments are unstable against rapid
collapse and cooling of the gas and probably form stars at their
centers.

In such a hierarchical picture (cf. Lake 1988) metal absorption
systems  at high redshift are more likely to arise from groups of 
relatively small, merging  protogalactic objects, rather than from 
ensembles of clouds in huge virialized halos:

(1)     Recent Keck spectroscopy of \op forest systems (Cowie et al.
	1995, Tytler et al.~1995) has shown that contamination
	of \op forest clouds by carbon is common in high-redshift \op
	absorbers with \hi\ column densities as low as a $N(\hi) \sim$ a
	few $\times 10^{14}$cm$^{-2}$. In the numerical simulations 
        such column densities typically correspond to gas densities
        smaller than those expected for
        fully  collapsed objects at these redshifts (indicating  
        baryonic overdensities of order ten), so at least some metal 
        absorption systems appear to occur outside virialized regions.

(2)     Typical observed temperatures of \civ\ systems are somewhat larger
	than expected if the gas were heated only by photoionization
	(Rauch et al.~1996, hereafter RSWB). Such an enhancement is
	predicted by numerical simulations (Haehnelt, Steinmetz \& Rauch
	1996, hereafter paper I) and is most likely due to
	shock and (to a smaller extent) adiabatic heating during 
        the gravitational compression of the gas. Absorber sizes along
        the LOS inferred from ionization balance calculations are of 
        order ten kpc. 	This is uncomfortably large for a 
        cloudlet-in-halo model (Haehnelt, Rauch, \& Steinmetz 1996, 
        hereafter paper II).

(3)     Simple scaling laws  predict that in a hierarchical scenario
	the typical ratio of cooling time to dynamical time
	decreases  as $(1+z)^{-3/2}$. 
	At high redshift the gas will generally cool  rapidly out to
	the virial radius.  It will be difficult to maintain large,
        massive, hot halos  for an extended period of time.
	Similarly for fixed circular velocity both the mass and the 
	size of the dark matter component of typical objects at high 
        redshift decrease as $(1+z)^{-3/2}$ (see Kauffmann (1996) 
        and Steinmetz (1996a) for quantitative  predictions).

\medskip

In  paper I we  used numerical hydrodynamical simulations to
demonstrate  that observed \civ\ absorption systems can be
qualitatively understood in terms of absorption by protogalactic clumps
(PGCs) formed through gravitational instability in a hierarchical
cosmogony.  The \civ\ and \hi\ absorption features caused by groups of
PGCs were found to resemble observed \civ\ and \hi\ absorption systems
if an overall homogeneous abundance of  [C/H] $\sim$ $-3$ to $-2$ was
assumed. Here, we will perform a quantitative analysis and extend the
work  to a larger set of diagnostically useful ionic species.  Section
2 will give the details of the numerical modelling.  In section 3 we
discuss the line formation process and the physical nature of the
absorbing structures in regions of ongoing galaxy formation. In section
4 we investigate  the observational consequences.  Conclusions are drawn
in section 5.

\section{Numerical simulations}

\subsection{The code}

The simulations were performed using GRAPESPH (Steinmetz 1996b).
This code combines a direct summation N-body integrator
with the Smoothed Particle Hydrodynamics technique
(Lucy 1977, Gingold \& Monaghan 1977) and is implemented for the 
special  purpose hardware GRAPE (Sugimoto et al.~ 1990).
The version of the code we use here is especially adapted to follow a
mixture of collisionless (dark matter) and collisional (gas) fluids. It
is fully Lagrangian, three-dimensional, and highly adaptive in 
space and time as the result of the use of individual smoothing
lengths and individual timesteps. The physical processes included in 
this version of the code include self-gravity, pressure gradients, 
hydrodynamical shocks, radiative and Compton cooling and photoheating
by a UV background with a specified spectrum. We do not assume rate 
equilibrium for the dominant baryonic species 
(H, H$^{+}$, He, He$^{+}$, He$^{++}$, and $e^-$) but
follow self-consistently their non-equilibrium time evolution. 
However, this had only  modest effect on the gas temperature in the 
lowest density regions. The non-equilibrium time evolution 
can generally be neglected for the questions addressed in this paper
but will be important for low column density systems at epochs close 
to reionization. We assume that the gas remains optically thin 
throughout the calculation and that the background radiation is
uniform  in space. Radiative transfer effects have been neglected 
(see also Navarro \& Steinmetz 1996).

\subsection{Initial conditions}

The initial conditions are identical to those in Navarro and Steinmetz (1996)
and were originally designed to study the formation of galaxies with circular
velocities between 80\,\kms and 200\,\kms. The background cosmogony
is a $\Omega=1$, $H_0=50\,$\kms\,Mpc$^{-1}$ cold dark matter (CDM)
universe with a
normalization of $\sigma_{8}=0.63$. The baryon fraction is $\Omega_b=0.05$.
Based on a P3M large scale structure simulation in a 30 Mpc box, eight
halos were selected (four each with circular velocity of about 
100\,\kms\ and 200\,\kms) and resimulated at much higher 
resolution. The tidal fields of the large scale matter  distribution
of  the original simulation were still included.
The high-resolution sphere has  a radius
 of about 2.3 (3.8) Mpc (comoving) for the
systems with a circular velocity of 100\,km/sec (200\,km/sec). The mass
per gas particle is $4.9 \times 10^6 M_{\odot}$ and $2.3 \times 10^7
M_{\odot}$ in the low- and high-mass systems, respectively.  We adopt
a Plummer gravitational softening of $2.5$ ($5$) kpc for the dark
matter and of $1.25$ ($2.5$) kpc for the gas particles in the low
(high) mass systems. All runs are started at $z=21$.

We use several different description for the intensity, redshift 
dependence and frequency dependence of the UV background. Most of 
the simulations were done using a power-law 
spectral energy distribution and the  redshift dependence suggested by Vedel, 
Hellsten \& Sommer--Larsen (1994).  We varied turn-on redshift,
spectral index ($\alpha=1\dots 2$) and normalization 
of the background ($J_{-22}=0.3 \dots 30$, in units of
$10^{-22}\,\mbox{erg}\,\mbox{s}^{-1}\,\mbox{cm}^{-2}\,\mbox{Hz}^{-1}\,
\mbox{sr}^{-1}$).   The more realistic description 
of the UV background proposed by Haardt \& Madau (1996) 
was also used. Although varying the UV background  had 
noticeable effects on the temperature and the ionization of the gas, 
the effects on the overall gas
distribution were small.  In this paper we will only briefly discuss these
effects and concentrate on simulations with $\alpha = 1.5$  and 
$J_{-22}=3$ for which the background was switched on at the 
beginning of the simulation ($z=21$). This corresponds to a typical 
quasar-like spectrum and is consistent with measurements of the helium
Gunn-Peterson decrement (Davidsen, Kriss \& Zeng 1996).  
The initial gas temperature was assumed to take the value, for which 
photoionization heating balances the cooling due to collisional 
processes and Compton scattering. 
For simplicity, we take the gas to be homogeneously contaminated by
metals (C,Si,N,O) with solar relative abundances and an absolute
metallicity of $10^{-2}$ solar.  Effects of time dependent and/or
inhomogeneous chemical enrichment were neglected. Later we shall
consider  different total and relative metallicities and briefly 
discuss implications of an inhomogeneous metal contamination.

\subsection{The matter and temperature distribution}

Some of the simulations were run up to the present epoch. 
For an individual  set of initial 
conditions, typically between one and three disk like galaxies with circular
velocities between 80 and 200\,\kms have formed by now. Each  
consists of several thousand particles. In addition, a couple of halos
with  smaller masses and circular velocities (about 50\,\kms ) are
present. Size, mass and circular velocities of these disks are in good
agreement with observed spiral galaxies. Compared to present-day 
spiral galaxies, however, the disks are too concentrated. This may 
be an artifact due to the neglect of energy and momentum input 
by star formation and active galactic nuclei.

\begin{figure}[t]
\centerline{
\psfig{file=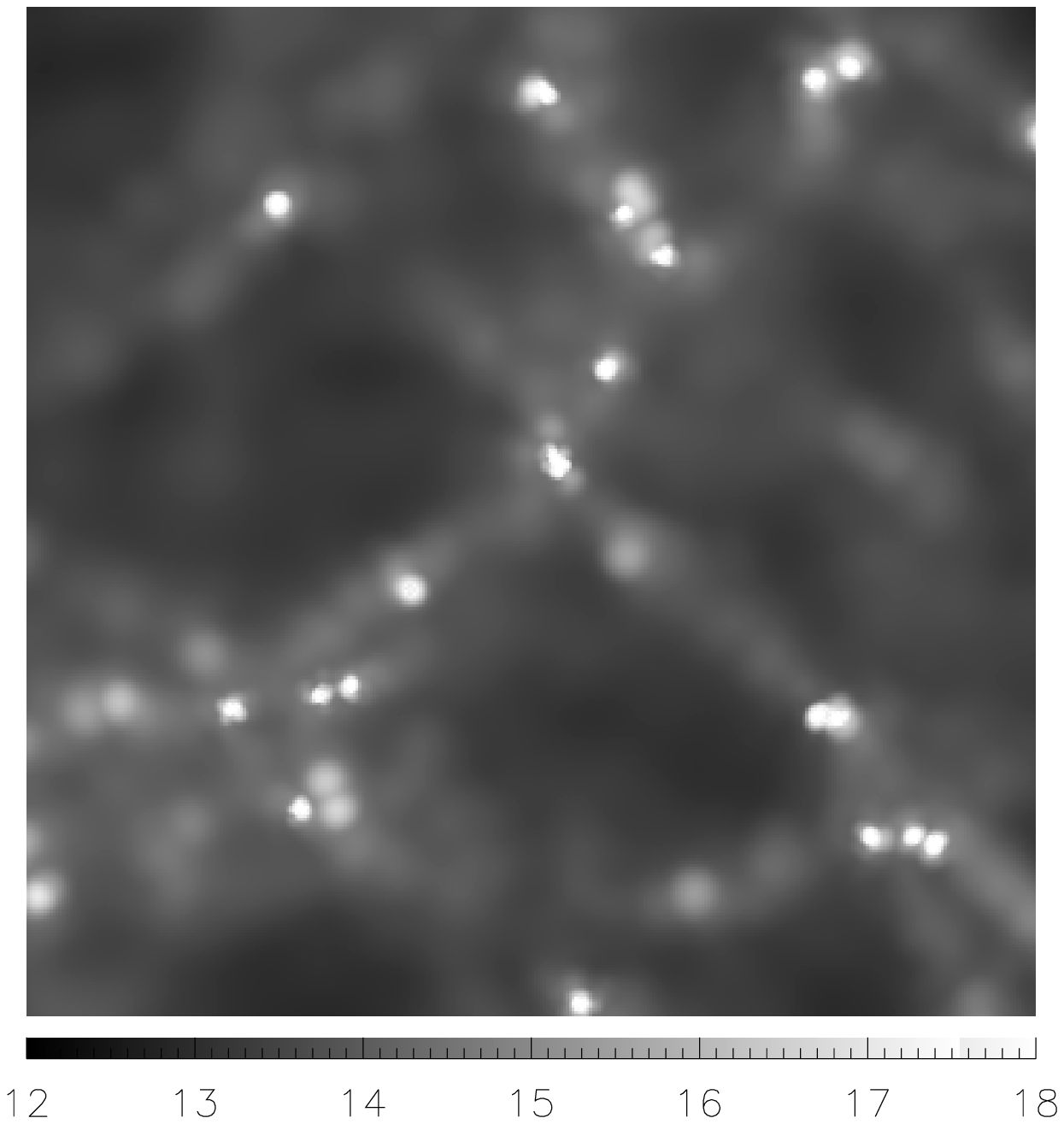,width=7.5cm,angle=0.}
\hspace{0.5cm}
\psfig{file=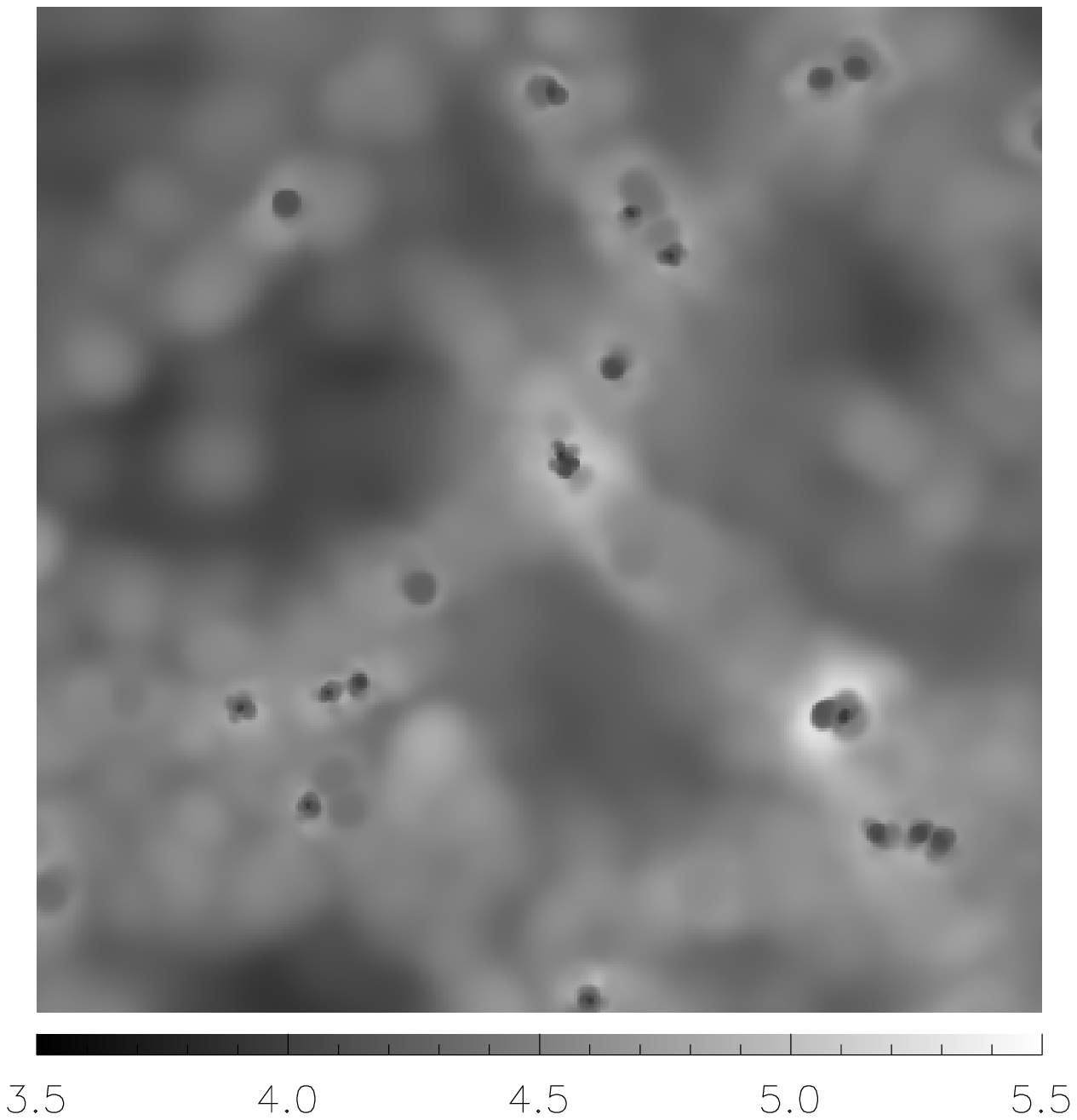,width=7.5cm,angle=0.}
}
\vspace{0.5cm}
\caption{\small The upper panel shows a grey-scale plot of the projected \hi\
column density ($\log N$)  of the inner  700 kpc of a simulation box at 
$z=3.07$ which will contain three $v_c \sim 100$\,\kms
galaxies at $z=0$.  The lower panel shows the 
temperature ($\log T$, \hi\ column density weighted) for the same box.
\label{greyscale}}
\end{figure}

Tracing the evolution of a present-day galaxy
backwards in time, one finds that at redshift three it has 
``split'' into  several  progenitors separated by  several 
hundred kpc. Generally the circular velocities of these progenitors are 
slightly smaller than that of the galaxy which they 
will later form, but  this is not true for  the most massive
progenitors. These have the same or even 
higher circular velocities even at redshifts  of three to five, although 
their mass may be a factor ten smaller (Steinmetz 1996a).
The progenitors probably suffer from the same overcooling problem 
as their present-day counterparts. This will have small effects 
for intermediate column density absorbers 
but conclusions regarding damped  \op absorbers should be drawn 
with some caution.

Figure \ref{greyscale}  shows the  column density  and
temperature distribution (\hi\ weighted along the LOS) at $z=3.07$ 
(bright=high column
density/hot) of a typical  simulation box in which three galaxies with
circular velocity $v_c \sim$ 100\,\kms\ will have formed at redshift
zero.  The area shown is 700\,kpc across (proper length). Individual
PGCs are embedded in a filamentary network.
Regions leading to a single $v_c\sim 200$\,\kms\ galaxy
look similar, with individual components scattered over
several hundred kpc.

The temperature in the gas rises from less than $10^4$\,K at the center of
expansion-cooled voids (large diffuse dark areas) to a few times 10$^4$\, K for
the gas in the filaments.  Higher temperatures up to several $10^5\,$K
degrees occur in spheroidal envelope within $\sim$ 30 kpc of the PGCs. These
envelopes of hot gas arise when infalling gas is shock-heated to temperatures
of the order of the virial temperature, but the density of this gas is low enough
that cooling times are long compared to dynamical time scales. The 
temperature, the location and the mass fraction of this hot gas component 
can depend sensitively on the assumed spectral shape of the UV
background and on the metal
enrichment (Navarro \& Steinmetz 1996). In the innermost few kpc
of the PGCs the cooling timescales are always short compared to the
dynamical time scale and the dense gas cools precipitously to 
temperatures below $10^4$ K.

\subsection{The ionization state of the gas}

\begin{figure}[t]
\centerline{
\psfig{file=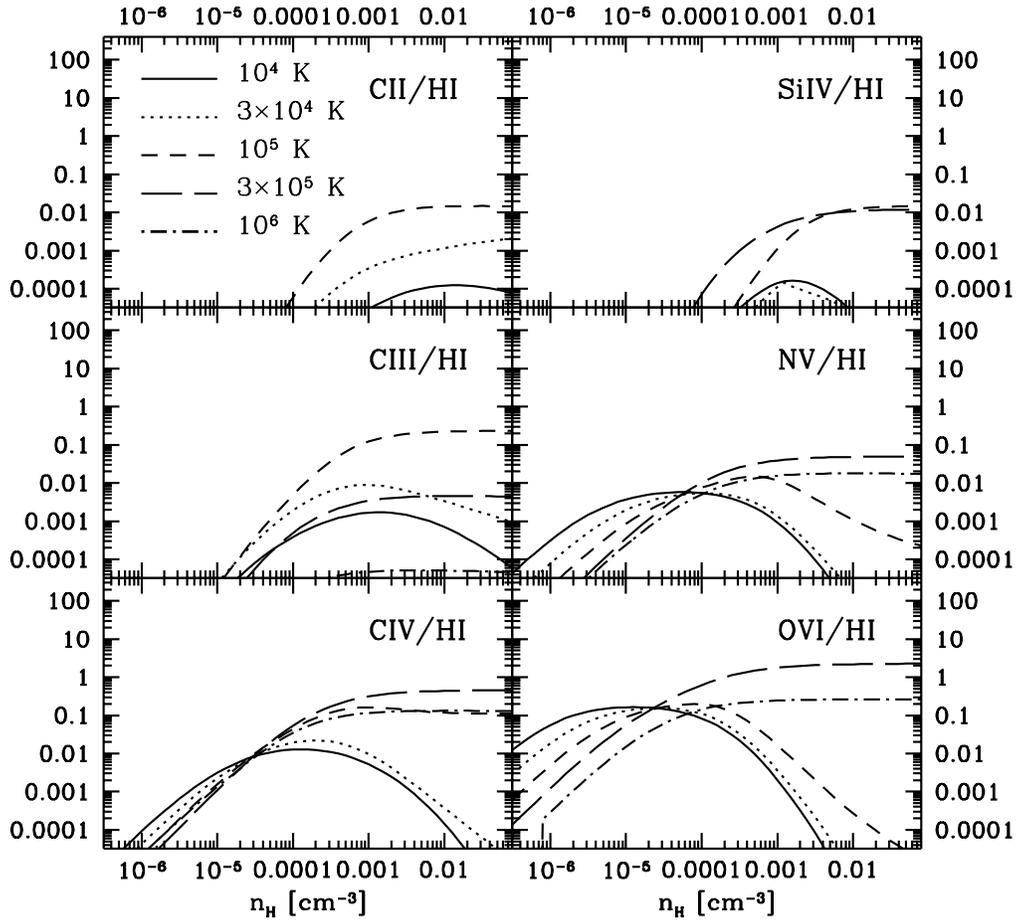,width=16.0cm,angle=0.}
}
\caption{\small Column density ratios relative to \hi\ for six different ionic
species calculated with CLOUDY for fixed density and
fixed temperatures between $10^4$\,K and $10^{6}$\,K 
as indicated on the plot.  A power law 
with $\alpha = -1.5$ and $J_{-22}=3$ was assumed for the background UV field.
\label{martin_1}}
\end{figure}

We have used the photoionization code CLOUDY (Ferland 1993) to
calculate the ionization state of the gas  for the  temperature,
density, and  UV background of each SPH particle in the simulation.
We assumed an infinite slab of gas of low metallicity, optically thin to
ionizing radiation and illuminated from both sides by a homogeneous UV
field.  (Self-)shielding from UV radiation in optically thick regions
was generally not taken into account.  Thus quantitative results for
LOS with column density above $10^{17} \cm^{-2}$ need to be viewed
with some caution. 
Below we will concentrate on the ionic 
species \hi, \cii, \civ, \siiv, \nv,
\ovi\ which are known to show the strongest observable lines in low and
intermediate column density absorbers.  As discussed in papers I and II,
observations and SPH simulations indicate that gas temperatures can
significantly deviate from the equilibrium value where photo-heating
balances line cooling processes. It is then important to use the actual gas
temperature to calculate the ionization state of the different states.
Figure \ref{martin_1} shows the ratios of various metal to \hi\ column
densities as a function of density for our chosen ionic species and a
set of temperatures between $10^{4}$ and $10^{6}$\,K .
At high and intermediate densities the metal column density  
generally increases relative to the \hi\ column density with
increasing temperature while at
small densities the opposite is true. \siiv\ and \civ\ are especially
temperature sensitive in the temperature and density range most
relevant for intermediate column density absorption systems around
$n_{\rm H} \sim 10^{-4} \cm^{-3}$ and $T\sim 5 \times 10^{4}$\,K.

\begin{figure}[t]
\vskip -2.0cm
\centerline{
\psfig{file=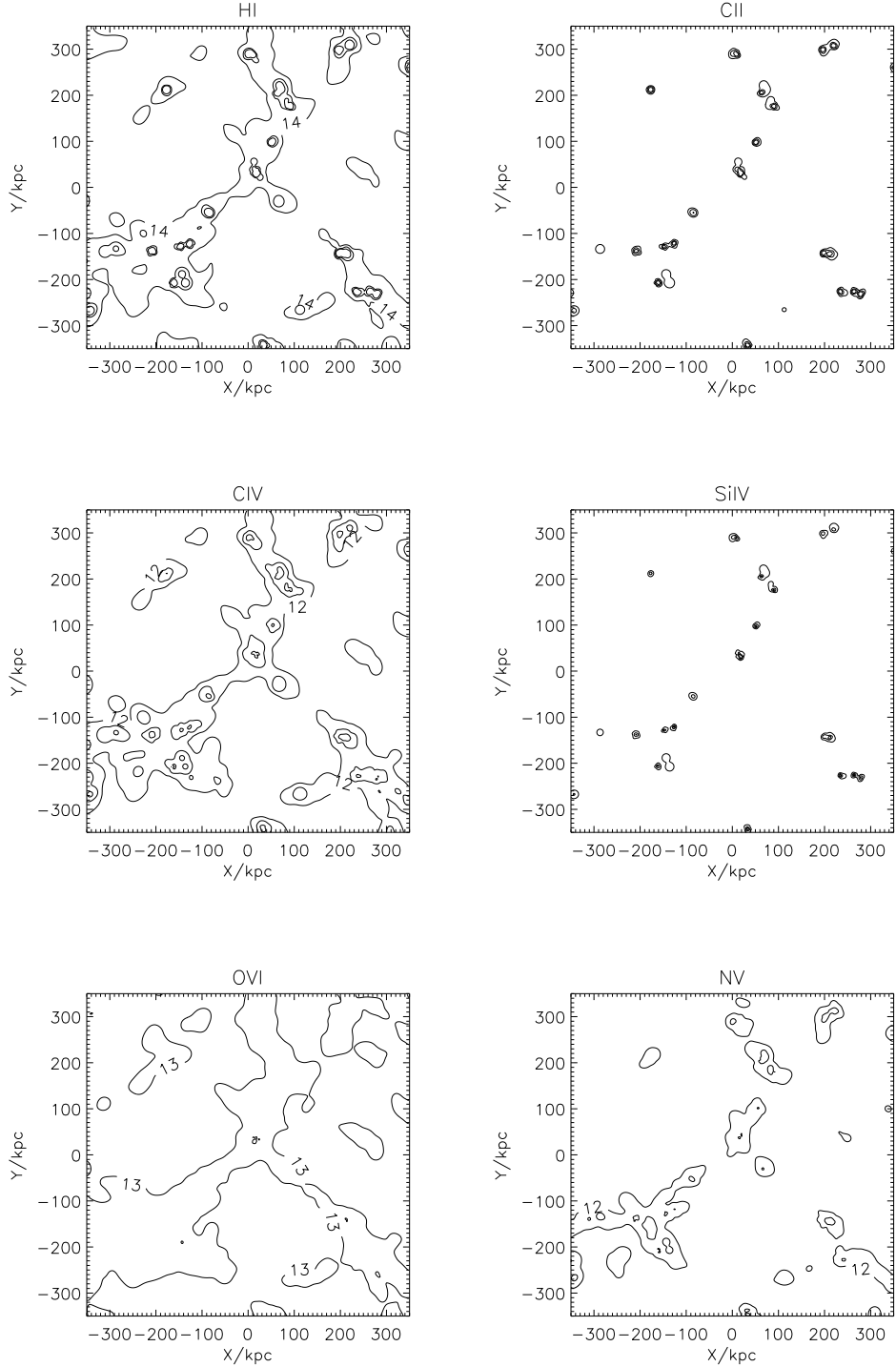,width=15.0cm,angle=0.}
}
\vskip -1.0cm
\caption{\small Projected column density of the inner  700 kpc of a
simulation box at $z=3.07$ which will contain three
$v_c \sim 100$\,\kms galaxies at $z=0$. Shown are
logarithmic column density contours in steps of 1 dex
for  \hi, \cii, \civ, \siiv, \nv\ and \ovi. [Z/H] = -2 and solar 
relative abundances were assumed.
\label{ioncontours}}
\end{figure}

\clearpage

\noindent

Figure \ref{ioncontours} shows  logarithmic column density contours for
all 6 ions in steps of 1 dex. \cii\ and \siiv\ are confined to dense regions
marking likely sites of star formation in the center of the PGCs.
\civ\ traces the hotter filaments and halos, while \ovi\ matches well
the low column density \hi\ contours and traces low density gas of any
temperature (cf. Chaffee et al.~1985, 1986).  In the box shown the
covering factor for detectable hydrogen absorption is 100\%.   The
same is true for \ovi\ if the metal contamination is indeed  homogeneous
and does not drop in regions outside the filamentary structure.
Approximately one third of all  LOS give detectable \civ\ absorption.

\subsection{Generating artificial QSO spectra}

To analyze the appearance of the galaxy forming region in absorption we
drew 1000 LOS and generated artificial spectra for the simulation box
shown in figure 1 (see section 2.3). 
The LOS had random orientations  and random offsets within $\pm
225$ kpc from the center of the box.  Optical depth $\tau(v)$ profiles
along the LOS were constructed projecting the Voigt absorption line
profile caused by the column density of each individual spatial pixel
onto its proper position in velocity space, using the relation (for
the \hi\ \op line)
\begin{eqnarray}
\tau(v) = \sum_{i}\tau_i(v-v_i) = 1.34\times 10^{-12} \frac{\dd N}{\dd v}(v)
\end{eqnarray}
(cf. Gunn \& Peterson 1965).
Here $\tau_i(v-v_i)$ is the optical depth of the Voigt profile at the observed velocity $v$, caused
by the column density $\Delta N_i$ in spatial pixel i, moving with
velocity $v_i=v_{\rm pec}(i)+v_{\rm Hubble}(i)$.
This can be expressed in terms of the column density per unit velocity,
$\dd N/\dd v$ at $v$. Units of $N$ are $\cm^{-2}$, $v$ is in $\kms$.

The spectra were made to resemble typical Keck data obtainable within a
few hours from a 16-17th magnitude QSO (S/N=50 per 0.043 \AA\ pixel, FWHM
= 8km/s). These ``data'' were then treated and analyzed in exactly the
same way (by fitting Voigt profiles) as actual observations.
An  absorption line was deemed detectable when the equivalent width
had a probability of less than 10$^{-5}$ (corresponding to a $>$ 4.75
$\sigma$ event) to have been caused by a statistical fluctuation.

\section{Absorption line formation in regions of ongoing gravitational
collapse}

Having access to the three-dimensional
gas/temperature distribution and the velocity field of the gas we are
able to study the line formation process as a function of the
characteristic physical properties of the absorbing structures.  We
expect the general mechanisms to be similar to those found earlier for
\hi\ absorption features (Cen et al.~1994; Zhang et al.~1995;
Hernquist et al.~1996; Miralda-Escud\'e et al.~1996). The emphasis of
the present work is, however, on the much narrower absorption lines
from metal ions. These  allow us to study the velocity structure  in 
galactic potential wells, where the corresponding \hi\ \op lines are
saturated and strongly blended. 
The resolution of the present simulations is much higher than that of
previous work, so we can pursue the physical quantities into
regions of larger densities, and study the fate of
objects of small mass.

\subsection{Environments causing specific absorption
patterns}

The following figures illustrate a few ``typical'' situations where
gravitational collapse gives rise to \civ\ and other metal absorption
features. The panels give (from top to bottom) the spectra of the six ions
offset vertically by 0.5 , the total baryon density, the temperature
and  the peculiar velocity. The $x$ axis is the spatial coordinate 
along the LOS labeled by the  relative velocity each position would 
have if following an undisturbed  Hubble flow. 
In the case of \civ, \siiv, \nv, and \ovi\  only the stronger
transition  of each doublet is shown. The dotted lines 
connect the spatial positions of overdense regions (selected manually)
to their density weighted positions in velocity space.

\subsubsection{Collapsing regions of moderate overdensity}

Figure 4a  shows a LOS  producing an intermediate column density
\hi\ profile. The line is saturated but not yet damped. In LOS passing very 
close to at least one PGC generally a single sharp density peak
dominates the 
absorption line formation (see the following pictures), but 
more often the LOS passes through regions of smaller overdensities 
just about to collapse into a single object. The peculiar velocity
field displays a typical infall pattern (redshift with respect to the 
Hubble flow for gas  falling in from the front,
blueshift for  gas falling in from behind, a jump at the location
of the density maximum). 

\civ\ and \ovi\ are the only metals which 
can be detected.  The absorption features are similar in appearance to the 
``partial \op limit systems'' studied by Cowie et al.~1995.
In this particular case the  individual density peaks have converged
in velocity space to form a caustic, but the enhanced density in physical
space is still much more important for  producing the absorption line
(see also  Miralda-Escud\'e et al.~1996).  In principle  velocity caustics
may  reduce the contribution of the Hubble expansion to the total
velocity width of absorption lines. We find, however, that this
rarely occurs. In most cases  infall velocities overcompensate
the Hubble flow. The ordering of two absorption lines in velocity
space can even be reversed relative to their spatial positions.

\subsubsection{Large velocity spread and chance alignments with
filaments}

Absorption caused by a chance alignment between the LOS and a
filamentary structure is seen in  figure 4b. The filaments 
generally lie at the boundaries of underdense regions (``voids'') 
and expand with velocities  in excess the Hubble flow. This is
apparent  from the divergent dotted lines and the ramp-like 
increase of the peculiar velocity as a function of real-space 
distance. Such structures rather than deep potential wells are 
likely to be responsible for the largest observed velocity 
splittings of \civ\ systems whereas moderate velocity splittings 
(up to 200 to 300\,\kms) are often due  to a complex 
temperature distribution or ionization structure  
internal to individual PGCs (see also  paper I).

\subsubsection{``Damped \op systems''}

Figure \ref{damped} shows two ``damped Lyman $\alpha$ systems''.
Currently we are limited to qualitative statements  about the
spatial extent and ionization state of  damped Lyman $\alpha$
systems as neither self-shielding nor energy/momentum input by
star-formation  or AGN  have been taken into account.  
Nevertheless, in agreement with Katz et al.~(1996)
we find that  for small impact parameters column densities 
becomes large  enough to produce damped \op absorption 
lines. 
Damped systems are usually dominated by one large density maximum. 
The density peak corresponds to a local temperature minimum due to 
cooling of the dense  gas at the center and is  surrounded by a shell 
of hotter infalling shock-heated gas. This leads to the characteristic
double-hump structure in the temperature diagram of figure 5b.  
The absorption lines for the various ions of the system shown in  
figure 5a appear rather similar to each other in terms of
position and line shape, with the exception of \ovi\ which arises 
at much larger radii.  In general, the maxima of the high ionization 
species for these highly optically thick systems need not coincide 
with the centers of the \hi\ lines and those of lower ionization 
species. Figure 5b  illustrates a situation where the component
structure has a very complex origin. While the strongest 
\cii\ and \siiv\ components coincide 
with the center of the damped \op line, the \civ\ and \nv\ positions are 
far off. This  indicates differential motion between the high and low
ionization regions.

\begin{figure}[t]
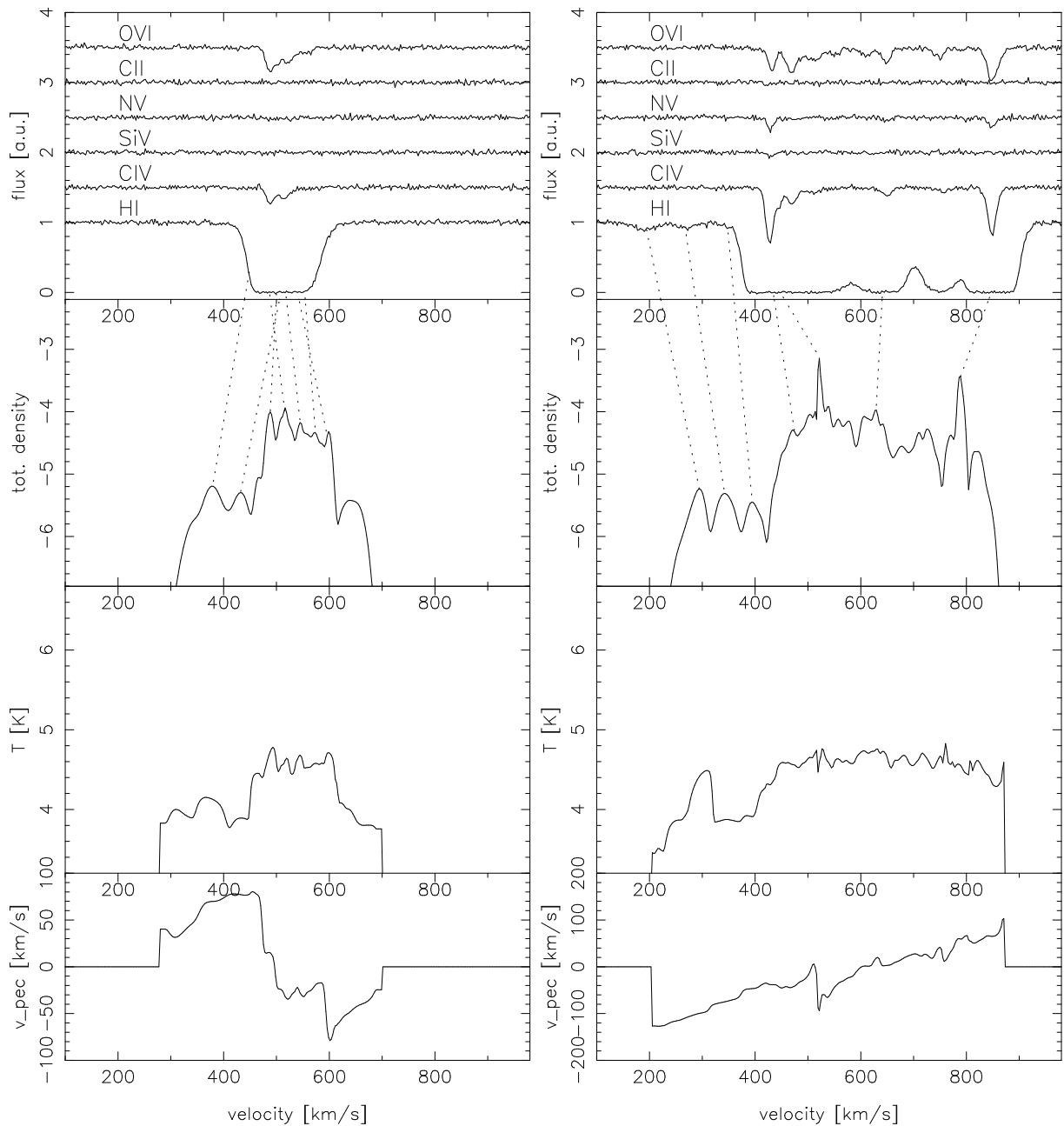

\vskip -1.0cm
\centerline{\psfig{file=haehnelt_4a.ps,width=8.cm,angle=0.}
\psfig{file=haehnelt_4b.ps,width=8.cm,angle=0.}
}
\vskip 1.0cm
\caption{\small left:  spectrum of a LOS through a collapsing region
of moderate overdensity  producing intermediate column density \hi\ and
weak \civ\ and \ovi\ absorption.  
right: LOS along a filament expanding 
faster than the Hubble flow.
The panels give (from top to bottom) the spectra of the six ions
offset vertically by 0.5, the total baryon
density, the temperature and the peculiar velocity (see section 3.1). 
In the case of \civ, \siiv, \nv, and \ovi\  only the stronger
component of each doublet is shown. 
\label{postshock}}
\end{figure}

\begin{figure}[t]
\vskip -1.0cm
\centerline{
\psfig{file=haehnelt_5a.ps,width=8.cm,angle=0.}
\psfig{file=haehnelt_5b.ps,width=8.cm,angle=0.}
}
\vskip 1.0cm
\caption{\small ``damped \op systems''.\label{damped}
}
\end{figure}

\clearpage

\begin{figure}[t]
\vskip -2.0cm
\centerline{
\psfig{file=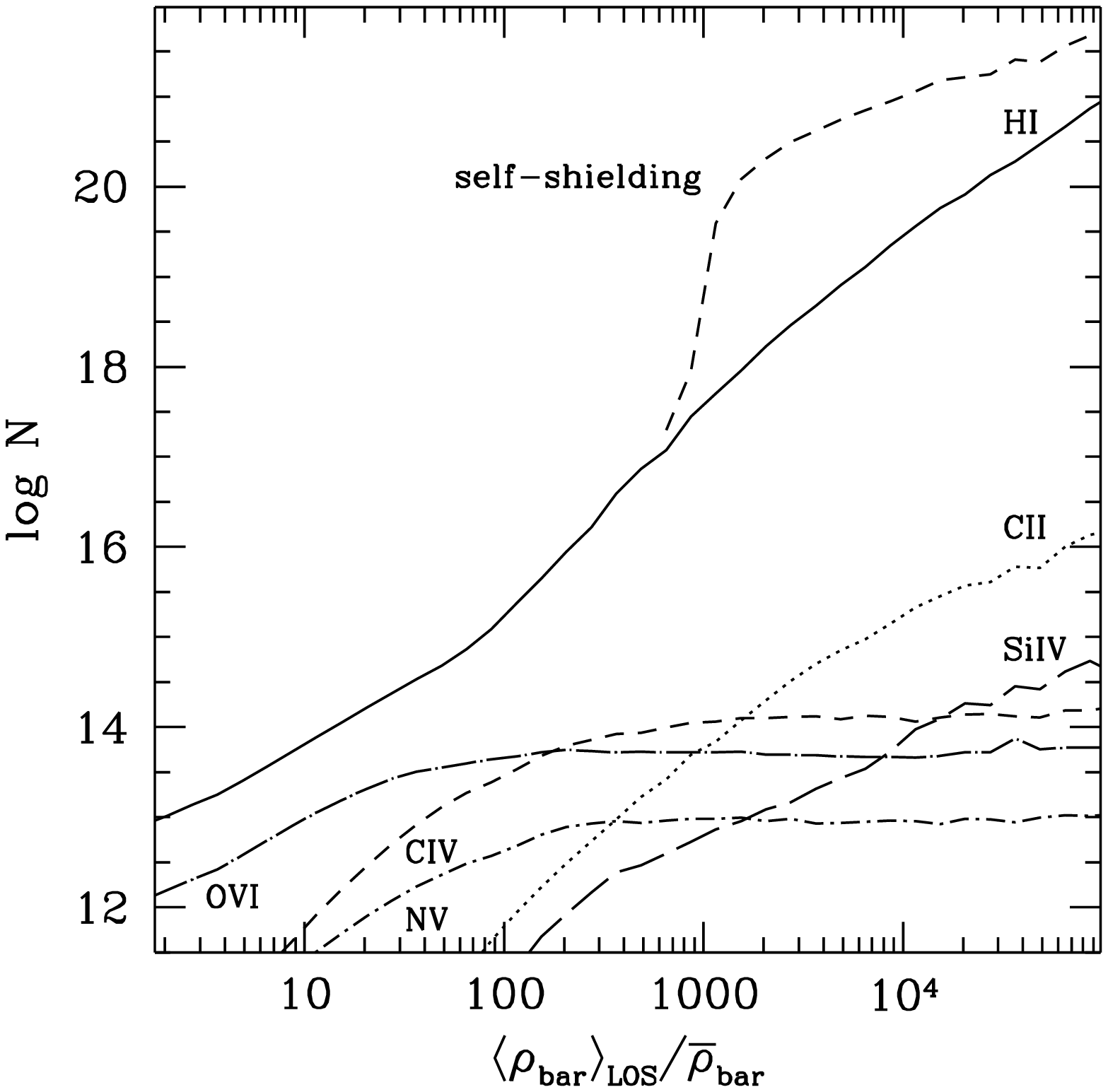,width=9.5cm,angle=0.}
\hspace{-1.5cm}
\psfig{file=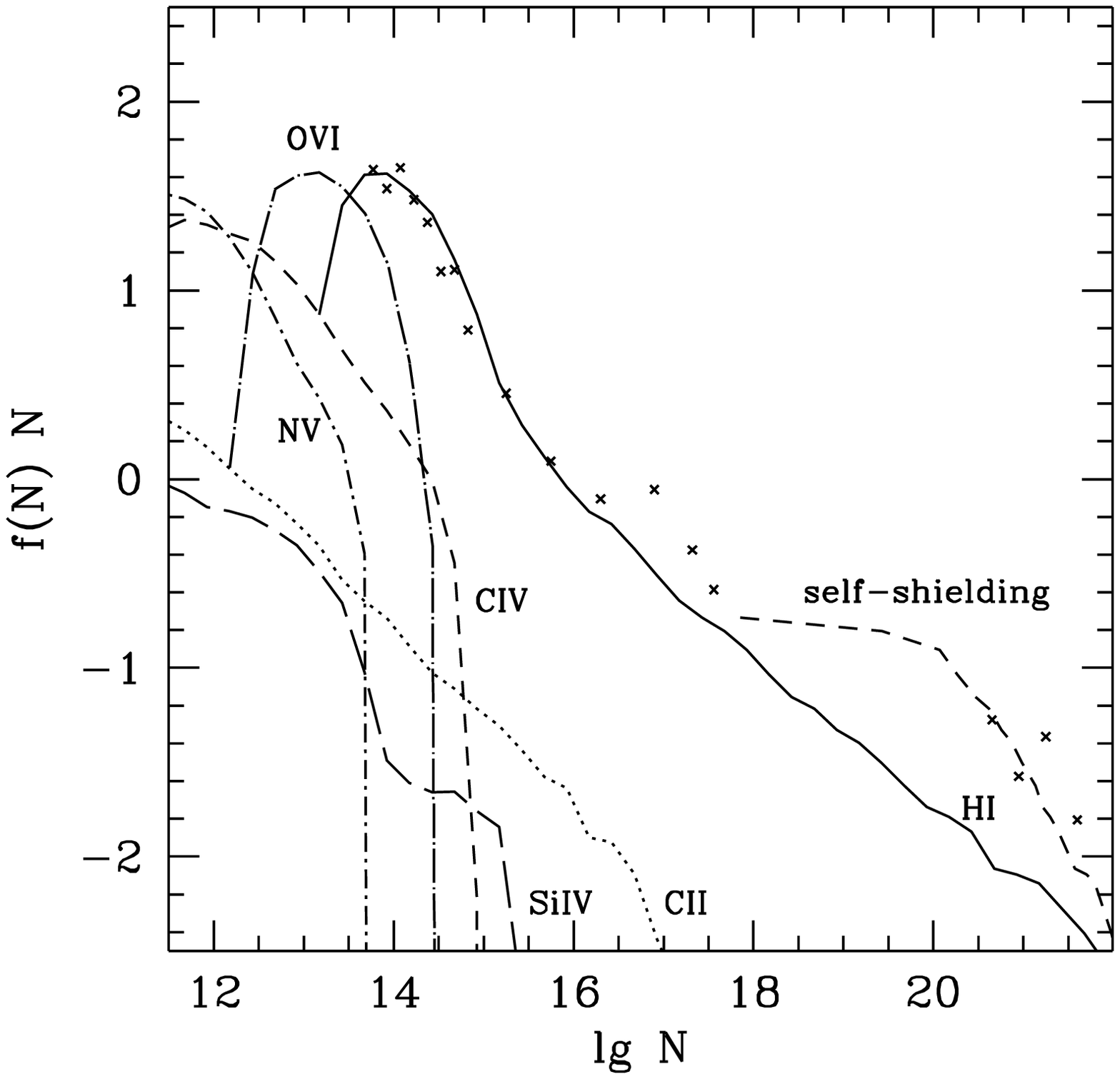,width=9.5cm,angle=0.}
}
\vskip -0.5cm
\caption{\small Left: mean projected column density
as a function of overdensity 
$\expec{\rho_{\rm bar}}_{\rm LOS}/\bar{\rho}_{\rm bar}$
(column density-weighted along the line-of-sight) 
for the simulation box shown in figure 1. 
Right: column density distribution (f(N) N) at $z=3$
for six different species in the simulation.  Crosses show the
observed column density distribution (Petitjean et al. 1993).
The column density normalization is described in section 4.1.}
\end{figure}

\subsection{The absorption properties of individual protogalactic clumps}

The solid curve in figure 6a shows the good correlation
between  the mean log $N$(\hi) and the baryonic overdensity in a typical
simulation box. Also shown is the mean log $N$ for our canonical set of
ionic species. As expected the strength relative to \hi\ varies
considerably from species to species. \cii\ and \siiv\ are strong at high
\hi\ column densities/baryonic over-densities, while \civ\ dominates at
intermediate column densities and \ovi\ probes the low density regime.
We have also plotted a simple self-shielding correction for  
large  \hi\ column densities. The correction was calculated 
with CLOUDY  specifying the total column density and mean density 
along the LOS.

\begin{figure}[t]
\centerline{
\psfig{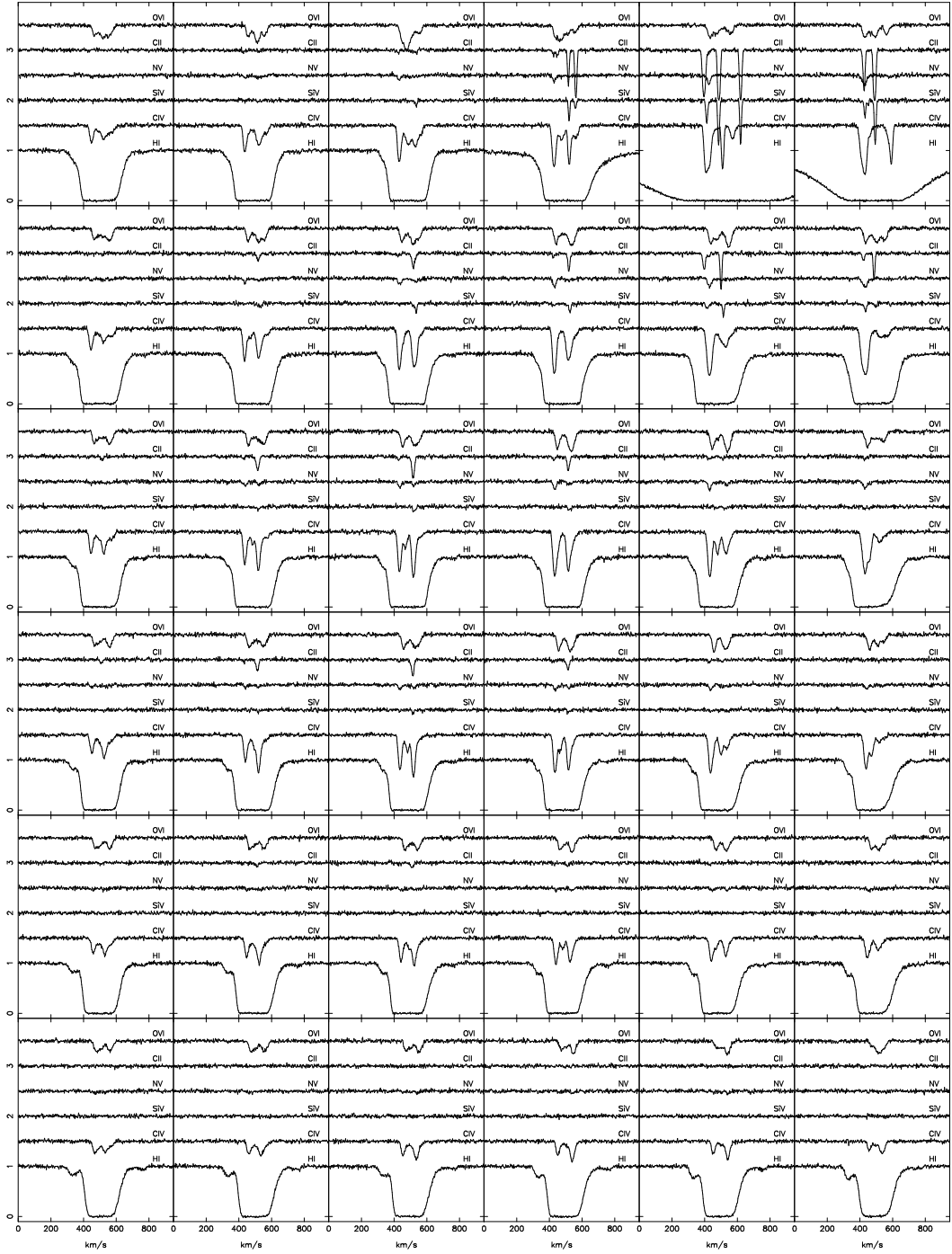}
}
\vskip 0.5cm
\caption{\small Mosaic of 6$\times$6 lines-of-sight near a collapsed
clump, separated by 3 kpc in each direction on the sky.
The center of the collapsed object is close to the top right corner.
The panels give the spectra of the six ions
offset vertically by 0.5 for clarity.
In the case of \civ, \siiv, \nv, and \ovi\ only the stronger line
of each doublet is shown.
\label{squ_6_3kpc_v}}
\end{figure}

Figure \ref{squ_6_3kpc_v} shows how the spectral features change with
distance from a fully collapsed clump. The plot consists of a mosaic of
LOS separated by 3 kpc from each other in the $x$ and $y$ directions on the
sky. The center of the clump is close to the top right corner.  
The \hi\ Ly$\alpha$ line exhibits damping wings within the
central $\approx$ 6 kpc.

\begin{figure}[t]
\vskip -1.0cm
\centerline{
\psfig{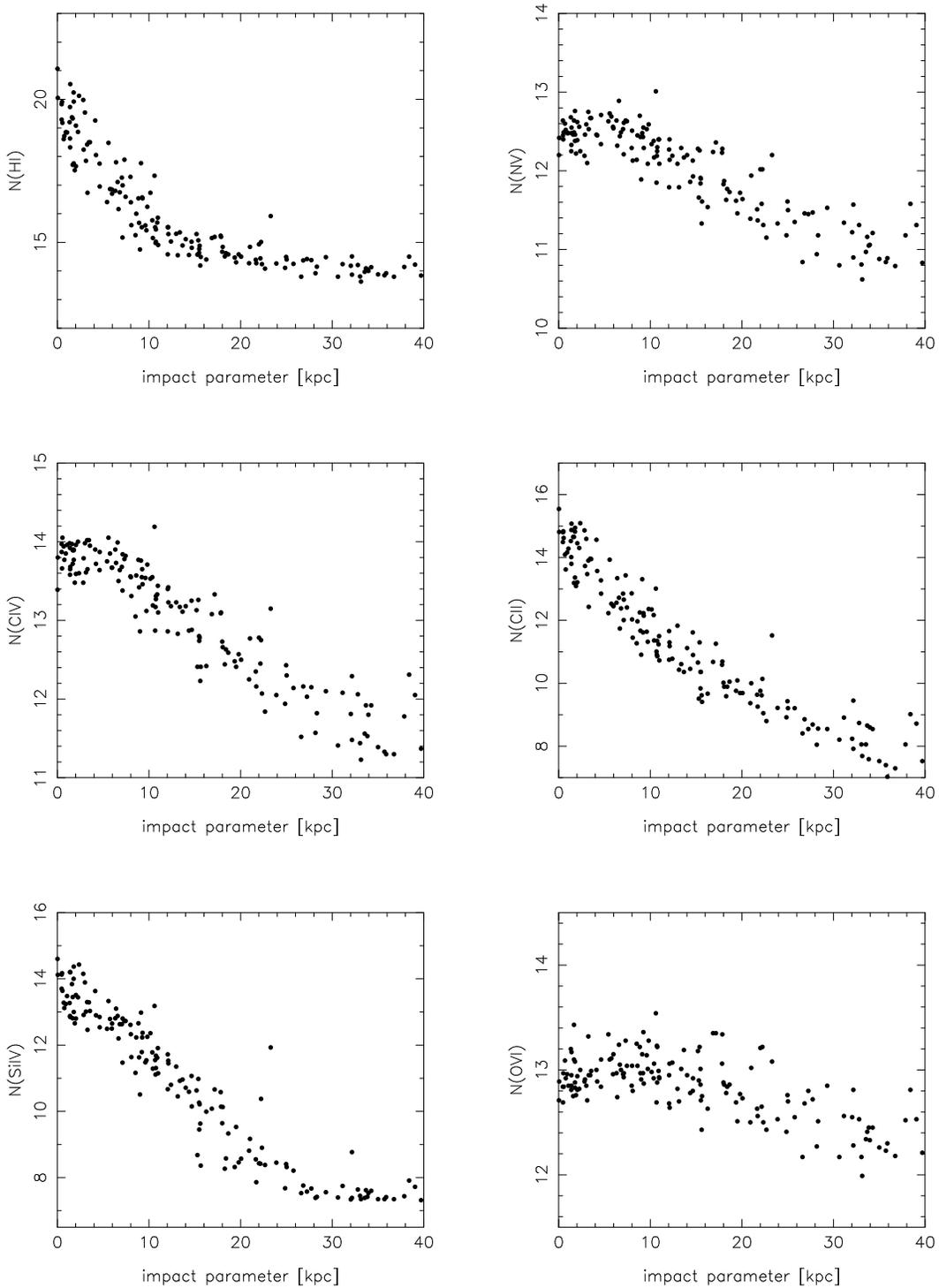}}
\vskip 0.5cm
\caption{\small Integrated column densities along random lines-of-sight
for the six ions as a function of impact parameter from the center
of a protogalactic clump with $1.2 \times 10^{9}$\,\Msol.\label{impaccol_ovi}}
\end{figure}

To quantify the spatial coherence of the absorption properties in more
detail we have  investigated a set of  random LOS close to a typical
PGC ($M \approx 1.2\times10^9 $M$_{\odot}$) in the simulation.  Figure
\ref{impaccol_ovi} shows the column densities as a function of impact
parameter for a set of randomly oriented LOS.  A sharp drop from $\log
N(\hi) \ge 20$ to $\log N(\hi) \approx 15$ occurs within the first five to
ten kpc.  At \hi\ column densities above $\log N \ga 17$
self-shielding should become important. This will lead to an 
even steeper rise 
toward smaller radii. After the rapid drop the \hi\ column density
decreases very gradually, still exceeding $\log N(\hi)\approx$ 14
at 100 kpc. The typical radius of the damped region of a protogalactic 
clump in our simulation taking self-shielding into account should be 
about five kpc. However, as pointed out previously  this value 
might increase once feedback processes are taken into account.

\begin{figure}[t]
\vskip -1.0cm
\centerline{
\psfig{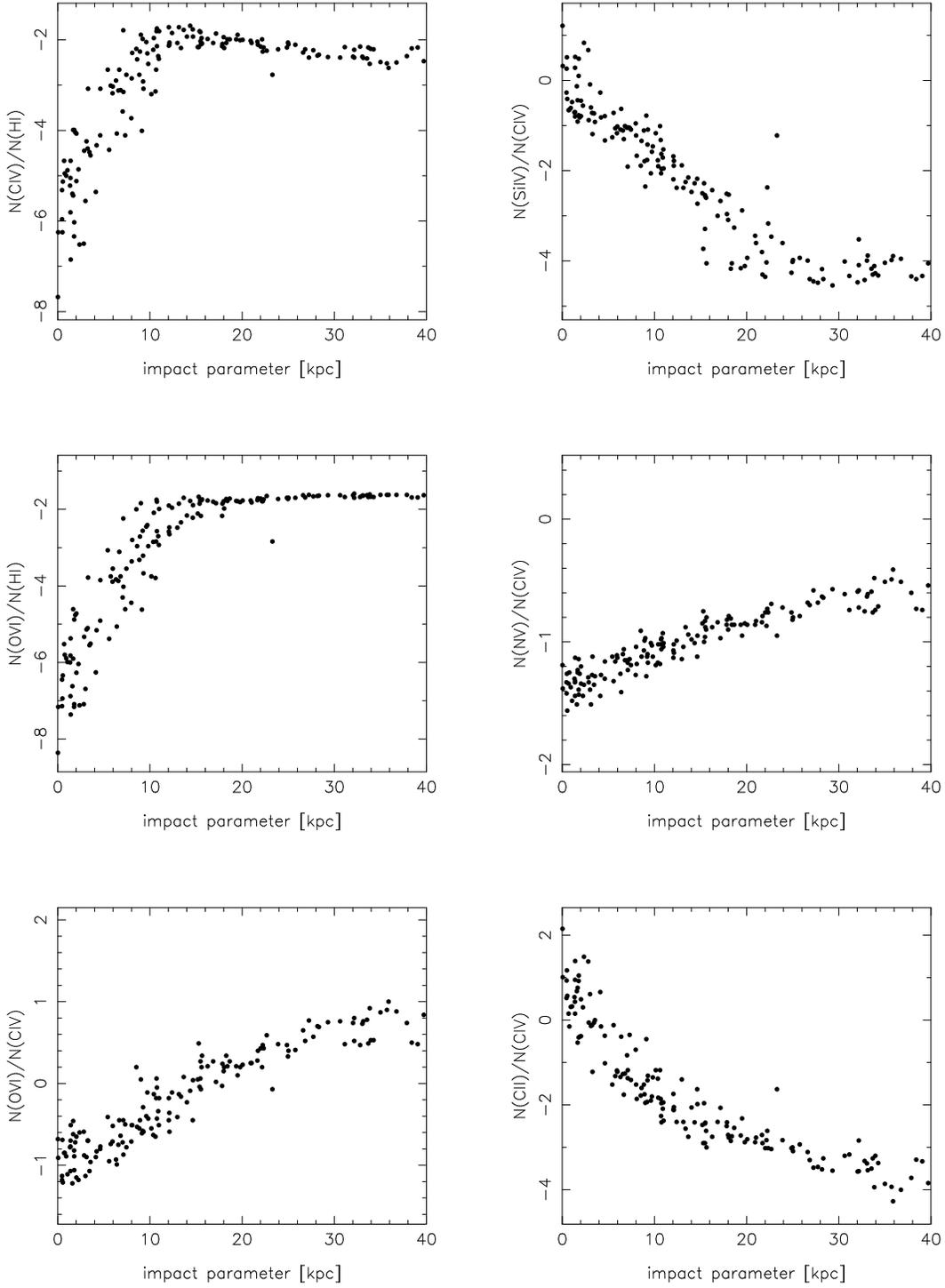}}
\vskip 0.5cm
\caption{\small Ratios of the integrated column density along random
lines-of-sight for several ions as a function of impact
parameter from the center of a protogalactic clump with 
$1.2 \times 10^{9}\,$\Msol.\label{impacratios}}
\end{figure}

Currently  a 10m telescope can detect \hi\ column densities of order
$10^{12}$ cm$^{-2}$ which means that the detectable \hi\ radius 
extends far beyond 100 kpc.  The largest detectable metal 
absorption cross-section  is subtended by \civ\ and probably \ovi\ 
(second panel on the left); at a detection threshold 
of $N$(\civ)$\sim 10^{12}$ cm$^{-2}$ 
(realistic for very high signal-to-noise data), 
the radius is $\sim $ 30 kpc; \siiv\ at the same significance level 
(taking the larger oscillator strength into account)
would be detectable out to about 12 kpc.  The decrease of the higher
ions \civ\ and \nv\ at the lowest radii is due to the increase in density;
it should be even more pronounced if  radiation transfer were
implemented.  Figure \ref{impacratios} gives the ratios of the LOS
integrated column densities for several ions. In particular, the third
panel on the left shows once more the importance of the \ovi\ ion, which
should be detectable at the largest radii with similar or even higher
significance than the \civ\ lines.  Thus, in spite of the difficulty of
observing \ovi\ in the \op forest the \ovi/\civ\ column density ratio should
be the best measure of the impact parameter.

\section{Observational tests}

\subsection{Column density distributions as a test of the gas distribution
and metallicities}

Before discussing the overall column density distribution of \hi\ and the
metal ionic species in our simulations we stress again that our
simulation boxes were  not chosen randomly (see section 2.2).  At our
redshift of interest ($z=3$) the mean baryonic density of the simulation
boxes containing  low (high) circular velocities halos is a factor 
1.27 (2.97) larger than the assumed mean baryonic density of the universe
for the inner $700$ (proper) kpc  cube. Furthermore,  
\hi\ column densities projected across the simulation box are 
generally larger than a few times $10^{13}$cm$^{-2}$. 
Below we will concentrate on
properties of the four boxes containing low circular velocities halos
at $z=3$ for which the overdensity is moderate and which should be
closest to a fair sample.


\begin{figure}[t]
\vskip -1.0cm
\centerline{
\psfig{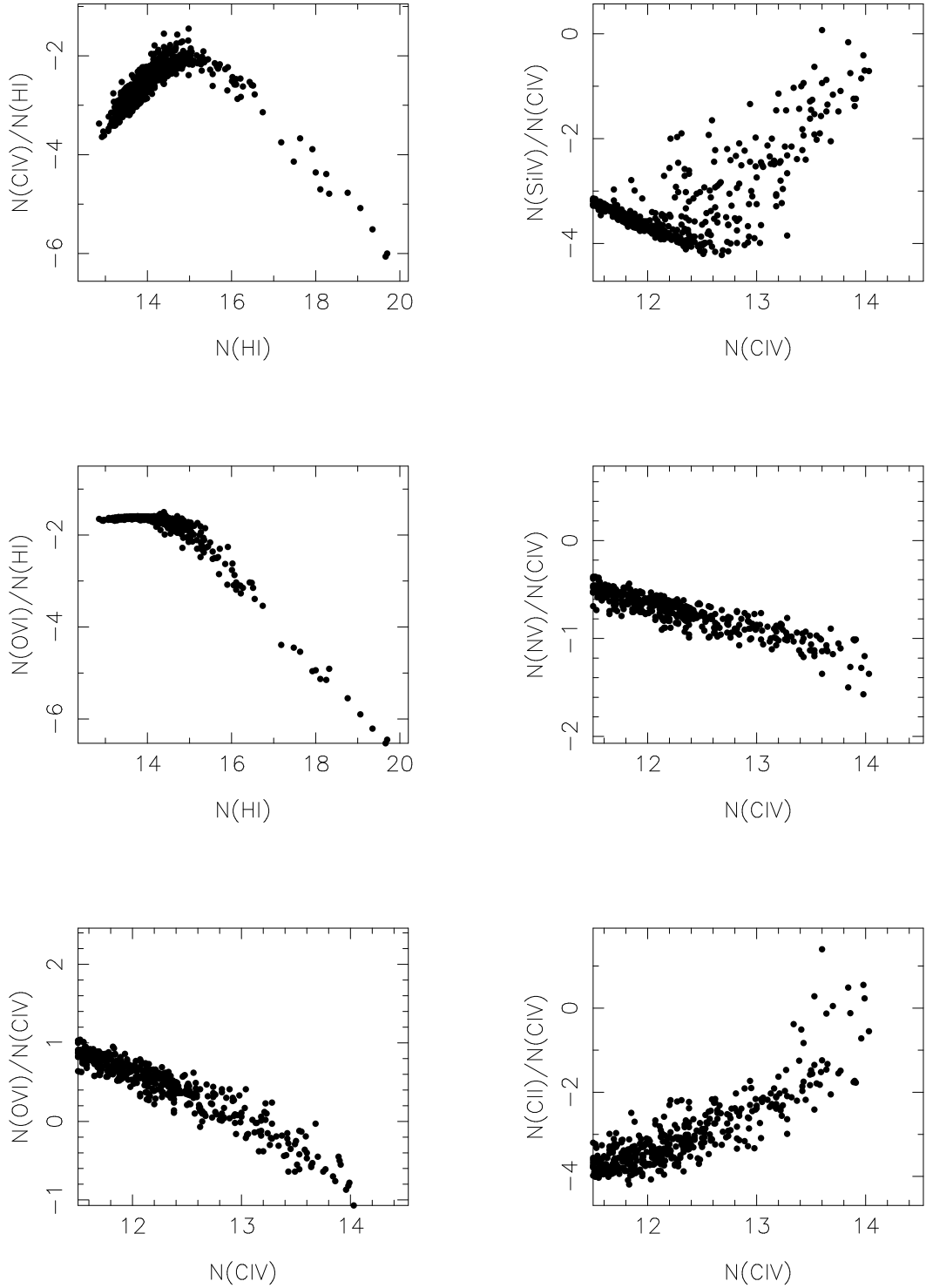}
}
\vskip 0.5 cm
\caption{\small Ratios of the integrated column densities along random
lines-of-sight for several ionic species as a function of \hi\
or \civ\ column density. \label{los_ionrat_ovi_999}}
\end{figure}

\clearpage

Column density distributions/ratios of ionic species with different
ionization potentials are an important diagnostic of the ionization
parameter, the spectral shape of the ionization field, the relative
element abundances and the temperature of the gas (e.g.  Bergeron \&
Stasinska 1986, Chaffee et al.~1986, Bechtold, Green \& York 1987,
Donahue \& Shull 1991, Steidel 1990, Viegas \& Gruenwald 1991, Ferland
1993, Savaglio et al.~1996, Songaila \& Cowie 1996). 
In our case the distributions of densities and temperatures are  
predicted by the simulation (only weakly dependent on the input 
radiation field) so the remaining adjustable parameters are the 
elemental abundances and the properties of the ionizing radiation.

\subsubsection{Column density distribution functions}
We define the differential column density distribution as usual 
\begin{equation}
f(N)   = \frac{\dd^{2} {\cal N}}{\dd X \dd N},
\end{equation}
where ${\cal N}$ is the number of lines and $\dd X = (1+z)^{1/2} \dd z$
(for $q_0$=0.5).
The curves in figure 6b  give $f(N) N$ at $z=3.07$ for the
four simulation boxes shifted in column density  by 0.3 dex, whereas the
crosses show the observed distribution (Petitjean et al.~1993). 
The data on the metal ions are again for a constant metallicity of
[Z/H]=$-2$. The self-shielding correction for large \hi\ column densities
mentioned in section 3.2 is also shown.  The  shape of the observed
distribution is matched reasonable  well by the simulations once the
self-shielding correction is applied.  As pointed out e.g. by Hernquist
et al.~(1996) and Miralda-Escud\'e et al.~(1996) the \hi\ column density
should scale with the baryon fraction, the strength of the ionizing
background and the Hubble constant as $\Omega_{b} ^{2} h^{3}/J$. 
With the parameters of our simulation the  applied shift of  0.3 dex 
in column density corresponds to 
$(\Omega_b\, h_{50}^{2}/0.05)^{2}\,/(J_{-22}\,h_{50}) \approx 1.5$, 
similar to the value found by  Hernquist  et al.~(1996). 
Taking into account that the mean baryonic density of the volume used
to calculate the column density distribution is a factor 1.27 larger 
then the  assumed overall mean baryonic density should somewhat 
lower this value. 
We do also find a deficit of systems with
\hi\ column densities  around  $10^{17} \cm^{-2}$, but it is smaller
than the discrepancy by a factor of ten reported by 
Katz et al.~(1996).

\subsubsection{Column density ratios for random lines of sight}

Figure \ref{los_ionrat_ovi_999} shows column density ratios for various 
ions,  integrated along randomly offset and oriented
LOS through the box described at the end of section 2.3.  The most
readily observable ratio is \civ/\hi, which in the simulations has a 
maximum of about $-2.2$ at $N$(\hi)=10$^{15} \cm^{-2}$ in the log.

\begin{figure}[t]
\vskip -1.0cm
\centerline{
\psfig{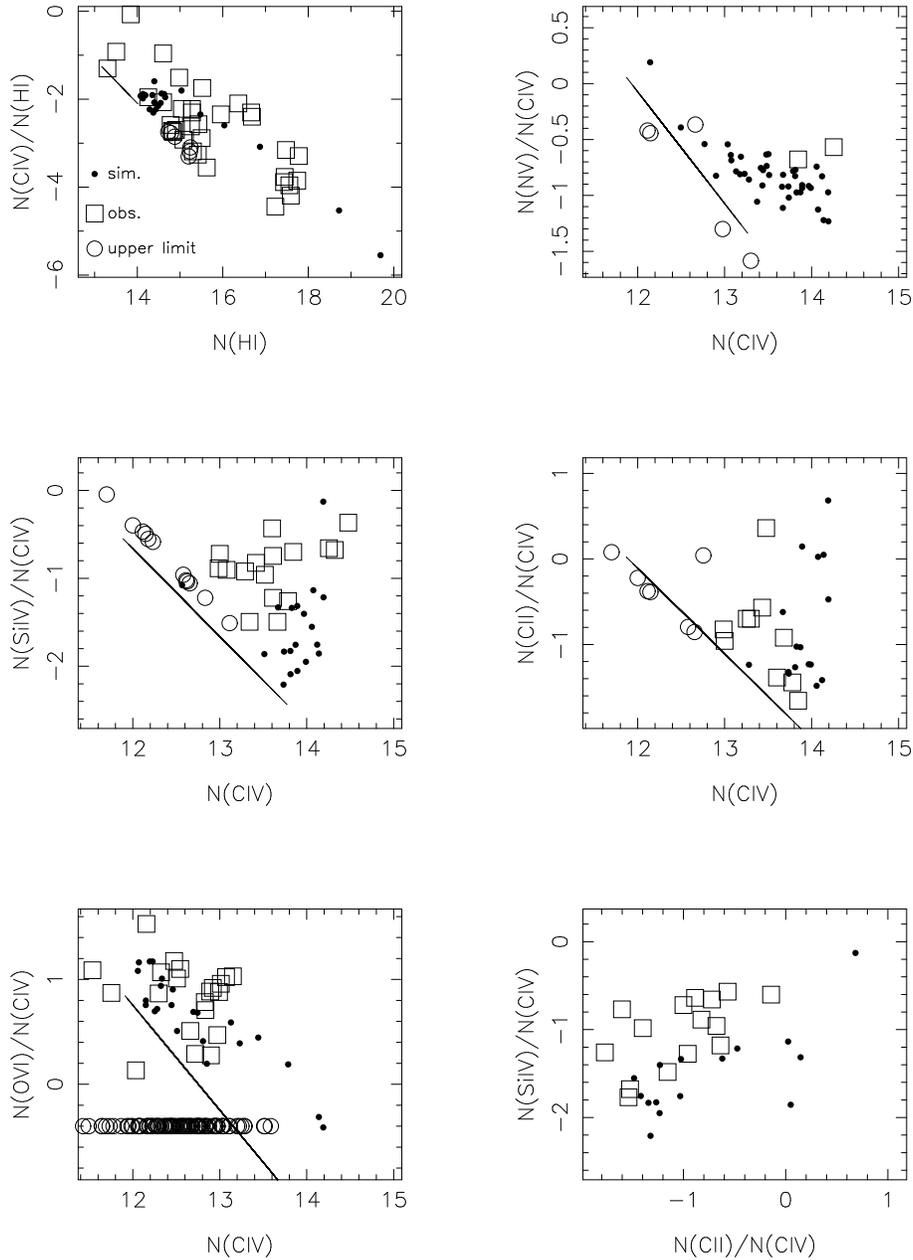}
}
\vskip 1.0cm
\caption{\small Column density ratios obtained from profile fitting the
simulated data. Absolute metallicity [Z/H]= $-2$, solar
relative abundances and a power law UV spectrum
with $\alpha$=$1.5$  and $J_{-22} =3$ were assumed.
Filled circles denote the ratios of the sums of the
column densities of all individual components along a LOS detectable at
the 4.75 $\sigma$ level.  The thin diagonal lines give upper limits
for the simulated lines in cases where only one ion (\hi\ in the first panel and \civ\ in the
other cases) has been detected. Open squares mark observed values (see
text), open circles either observational upper limits or (bottom left panel)
\ovi\ values undetermined because of blending.
\label{fitresults}}
\end{figure}

\clearpage 

\noindent 
This is consistent with the range of values measured by 
Cowie et al.~(1995), but a better
match can be found with a carbon abundance of  [C/H]= $-2.5$ instead
of  the originally adopted [C/H]= $-2$.  At higher column densities 
\civ\ recombines to form \ciii\ and \cii\ with increasing density and
shielding. At lower column densities \civ/\hi\ also declines as carbon
becomes more highly ionized into \cv. This effect will cause
difficulties for searches for carbon enrichment in the \op
forest at column densities much lower than the presently accessible
limits in \hi\ (10$^{14}$ cm$^{-2}$), as not only the \hi\ 
column but also the \civ/\hi\ ratio declines.

Results obtained from Voigt profile fitting a subset of the spectra
(using the software package VPFIT, Carswell et al.~1987) are shown in
figure \ref{fitresults}.  Solid dots are measurements from the
simulation.  For comparison open boxes give observational data
for \cii, \civ, \siiv\ and \nv\ by Songaila \& Cowie (1996) 
and for \ovi\ from Rauch et al.~(in prep.).
The circles show observational upper limits 
or indicate  undetermined \ovi\ in cases of severe
blending with \op forest lines. 
The thin diagonal lines indicate the  approximate
detection limits for absorption features in the 
simulated spectra. Most of the general trends in the
observed column density ratios are well reproduced. The scatter 
in the column density ratios between different metal ionic species
is similar to that found observationally. This suggests that the
simulations  produce a realistic range of ionization conditions.

There are, however, some interesting  discrepancies between simulated 
and actual data. First we note that the scatter in \civ/\hi\ (panel on
top left) for the observations is much larger than that in the
simulations (where constant metallicity was assumed). We take this to imply 
that the carbon fraction and probably also the absolute 
metallicity in the outskirts of protogalactic
objects at $z\sim$ 3 fluctuates over one to two  orders of magnitude
throughout the column density range $N$(\hi) $10^{14}$ to  $10^{18}$
cm$^{-2}$. It is interesting to note that a similar 
scatter is seen  in [C/Fe] in metal poor stars 
(McWilliam et al.~1995, Timmes, Woosley \& Weaver 1995).

There are also some obvious differences between the absolute
levels of the observed and simulated  mean column density ratios.
While the abundance ratios for the high ionization species (\civ, \nv, \ovi)
in the simulation agree reasonably well with the  observations, \siiv\
and \cii\ are significantly off, with the simulated \siiv/\civ\ and
\cii/\civ\ ratios lower by a factor ten than the observed values.

\subsubsection{Matching the observed column density ratios}

Some of these discrepancies can be reduced by an appropriate modification
of the relative abundances. So far we have assumed
{\it solar relative abundances}. The generally low abundances
in present-day galactic halo stars (which may
contain a fossil record of the high $z$ gas abundances) as well as
optically thin \civ\ systems (Cowie et al.~1995) and high-redshift  damped \op
systems (Pettini et al.~1994, Lu et al.~1996) indicate that
relative metal abundances like those in very metal poor
stars (McWilliam et al.~1995) may be 
more appropriate. Below we will discuss the effect of changing the  
abundance pattern from solar values to:  [C]=[N]=0; [Si]=[O]=0.4. 
The brackets denote the difference to solar abundances in dex.

\begin{figure}[t]
\vskip -1.0cm
\centerline{
\psfig{file=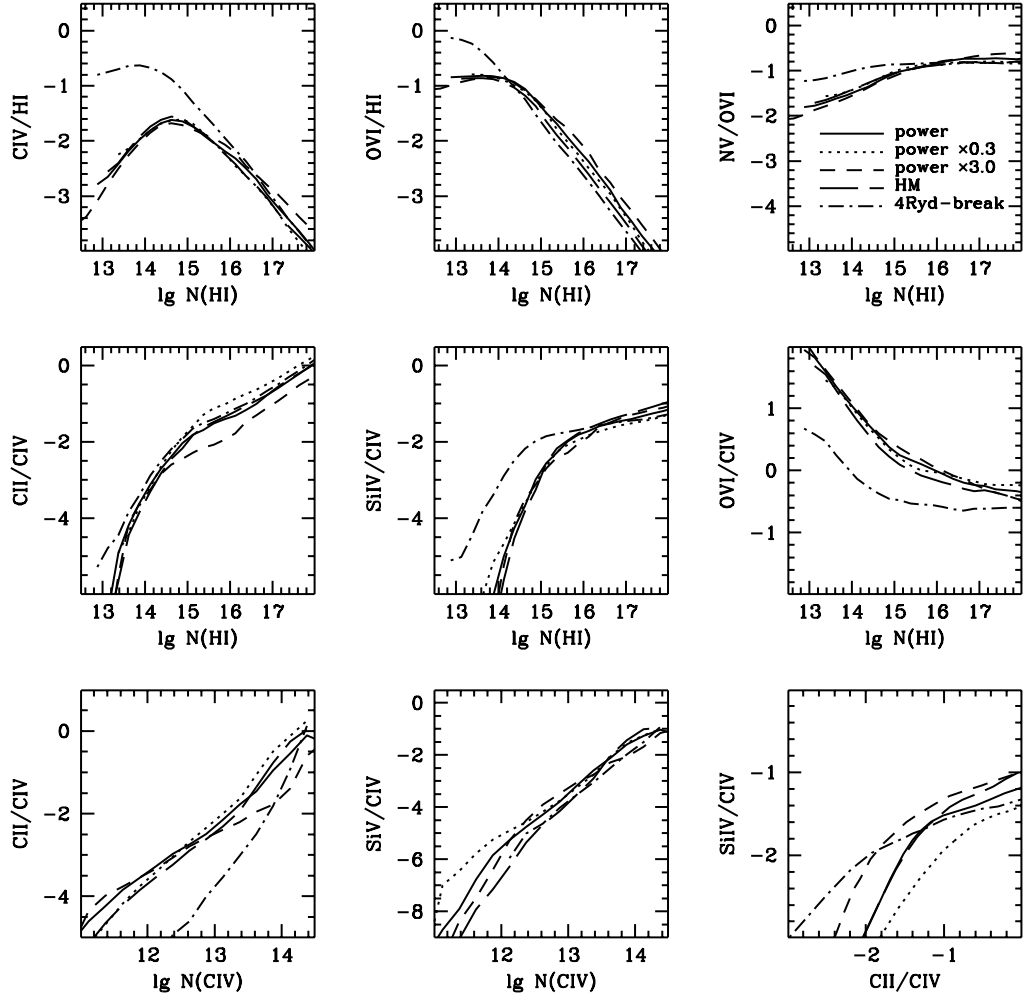,width=15.cm,angle=-0.}
}
\vskip 1.0cm
\caption{\small Column density ratios for various
UV background radiation fields. The solid curve are for  the 
$\alpha$ = 1.5 power law spectrum.
The other curves are for: a power law with
intensity increased and decreased by factors of 3;
a spectrum taking intergalactic absorption and emission
into account (Haardt \& Madau 1996); and  a power law with a
step function like break at 4 Rydberg  (flux reduced by a factor 100
at 4 Rydberg, constant flux for higher energies up to the point where
the flux equals that of the underlying power law). \label{martin_5}}
\end{figure}

Some improvement may also be expected from a different intensity
or spectral shape of the ionizing radiation background.
Figure \ref{martin_5}  shows the influence of such changes on various
column  density ratios.  The solid curve shows the values for 
the $\alpha$ = 1.5 power law spectrum.
The other curves are for several different spectral shapes and normalizations  
of the UV background:
a power law with intensity increased and 
decreased by factors of 3; a spectrum taking intergalactic 
absorption and emission into account (Haardt \& Madau 1996); and  
a power law with a step-shaped break at 4 Rydberg  
(flux reduced by a factor 100 at 4 Rydberg, constant flux for higher 
energies up to the point where the flux equals that of the underlying 
power law). The effect of the absorption edges in the Haardt \& Madau spectrum
on the ion ratios considered here is
generally small for the relevant column densities.
A power law with slope 1.5 is a good approximation.
Changing the normalization has noticeable but moderate effects on the
\siiv/\civ\ and \cii/\civ\ ratios. Otherwise the changes are again small.
The case of a powerlaw with a 4 Ryd cutoff, however, leads to strong
departures in the ion/\hi, and in the ion/ion ratios for a given \hi\ 
column density.  A spectrum of the last kind was suggested by
Songaila \& Cowie (1996) and Savaglio et al.~(1996) to
account for large observed \siiv/\civ\ ratios at $z>3$.  These authors
consider an increasing \heii\ opacity  or an  increasing stellar
contribution  to the UV background towards higher redshifts as
possible causes for a softer spectrum.

However,  for a fixed \civ\ column density the power law with
4 Rydberg  cutoff actually even lowers the \siiv/\civ\ and \cii/\civ\ 
ratios relative to those obtained with the originally 
adopted $\alpha$ = 1.5 power law.
Moreover, such a spectrum would imply a factor three to ten smaller
carbon abundance. Apparently, changing the UV radiation field does
not dramatically improve the overall agreement with the observed ion ratios.

\begin{figure}[t]
\vskip -1.0cm
\centerline{
\psfig{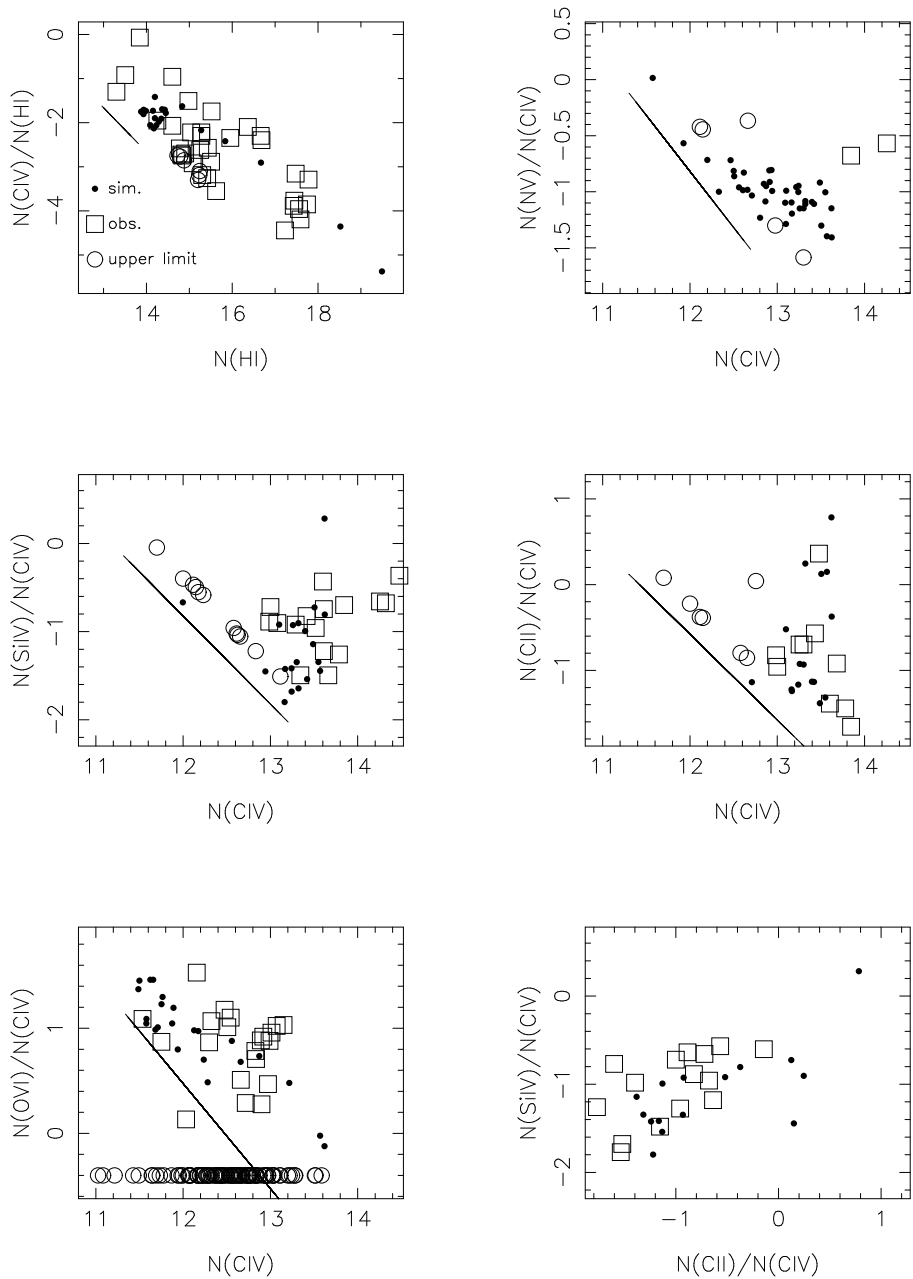}
}
\vskip 1.0cm
\caption{\small Column density ratios obtained from profile fitting the
simulated data (filled circles). Compared to figure 11
the absolute metallicity has been changed from [Z/H] = $-2$
to $-2.5$,  solar relative abundances have been replaced by those of
metal-poor stars ([C]=[N]=0, [Si]=[0]=0.4) and the spectrum
suggested by Haardt \& Madau (1996) is used instead of a power
law with $\alpha =1.5$.
The diagonal lines give upper limits for the simulated absorption features
in cases where only one ion (\hi\ in the first panel and \civ\ in the
others) has been detected. Open squares mark observed values (see
text), open circles either observational upper limits or undetermined 
\ovi\ .
\label{bestfit}}
\end{figure}

The observed and simulated column density
ratios are compared in figure \ref{bestfit} after 
adopting some of the changes discussed
above (metallicity  reduced by 0.5 dex to [Z/H] =$-2.5$;  metal poor
abundance pattern; spectrum  as suggested by Haardt \& Madau).  The
discrepancy for \siiv\ and \cii\ has now been significantly reduced: the
contribution to the shift in the \siiv/\civ\ vs. \civ\ came in equal parts
from the adjustment in relative and absolute metallicity, whereas the
\cii/\civ\ vs. \civ\ ratio improved mostly because of the decrease in
absolute metallicity.  Thus the agreement is now quite good for \siiv,
\cii\ and \civ. \ovi\ and \nv\ are slightly further off but --
given the large observational uncertainties in these two ions --
probably consistent with the data.

\clearpage

\begin{figure}[t]
\centerline{
\psfig{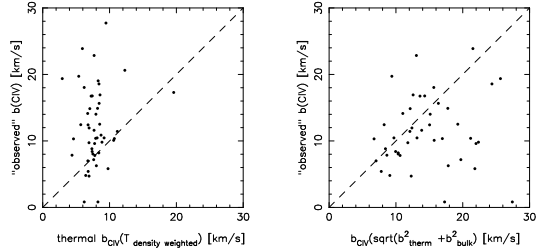}
}
\vskip 0.5cm
\caption{\small Left panel: Total \civ\ Doppler parameter {\it
measured} by profile fitting versus density weighted \civ\ thermal
Doppler parameter computed from the gas temperature in the
simulation. Right: Total  measured \civ\ Doppler parameter 
vs.~the Doppler parameter  obtained by adding  RMS bulk velocity 
dispersion and  thermal velocity in quadrature. 
\label{bmeasbconstr}}
\end{figure}

\subsection{The Doppler parameter as indicator
of gas temperatures and small-scale bulk motions}

Temperature and bulk velocity measurements from individual absorption lines
are useful discriminants
for the environment of heavy element absorption systems.  The actually
observable quantity is the Doppler parameter $b$ (=$\sqrt{2}\sigma$) of
the absorbing line. The line width results from a convolution of
the thermal motion and the small scale bulk motion.  Measurements of the
Doppler parameter for ionic species with different atomic weights permit
a decomposition into the contributions from thermal and non-thermal
(bulk) motion. The precise nature of the decomposition depends,
however, on the velocity distribution of the bulk motion.

For the simulations the relative importance of the
contributions to the line formation can be easily studied
as we have full knowledge of the density and peculiar velocity 
field along the LOS. Here we proceed as follows: first we
compute the column density weighted temperature and RMS velocity
dispersion in an overdense region selected manually; then we fit the
absorption line closest to the position of the velocity centroid of the
region with a Voigt profile and compare results.

\begin{figure}[t]
\centerline{
\psfig{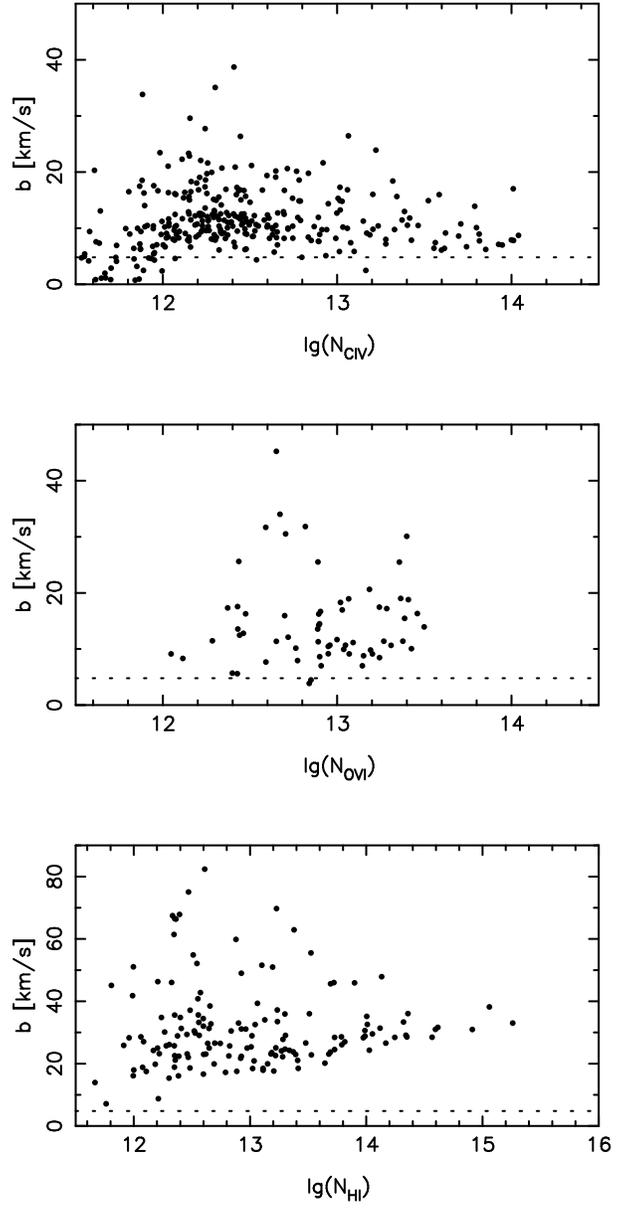}
}
\caption{\small Doppler parameter -- column density ($b$--log$N$)
diagrams for \civ\ (top), \ovi\ (middle)  and \hi\ (bottom).
The dotted lines give the
spectral velocity resolution.
Note the different scales of the axes.\label{bcivbovibhi}}
\end{figure}

\clearpage
\begin{figure}[t]
\centerline{
\psfig{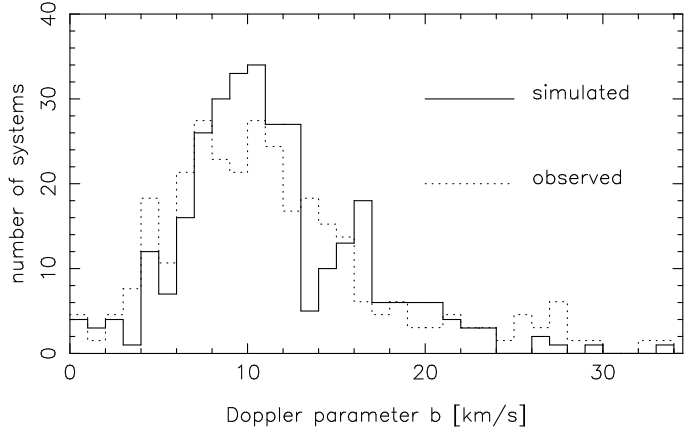}
}
\vskip 0.5cm
\caption{\small Comparison between the \civ\ Doppler parameters from
the simulation (solid line) and the observed distribution from
Rauch et al. 1996 (dotted line).  The observed distribution has been
scaled such as to match the total number of lines 
of the simulated distribution.\label{bsimobscomp}}
\end{figure}

\noindent
The top panel in
figure \ref{bmeasbconstr} shows a plot of the ``observed'' \civ\ 
Doppler parameter from the fit versus the
purely thermal Doppler parameter computed straight from the temperature
array of the simulation. While the density weighted thermal \civ\ Doppler
parameter of many LOS hovers around 7 to 8 $\kms$, the ``observed''
values occur over a wide range. The total \civ\ Doppler parameter is
obviously not an unbiased measure of the temperature alone. There
are substantial varying contributions from bulk motion.  It is not
obvious how  temperature and bulk motion should be deconvolved.
Making the simplest possible ansatz of adding them in quadrature (which would be strictly
true if the bulk motion followed a Gaussian velocity distribution) we
arrive at the results shown in the lower panel:   the observed line
width does indeed measure the quadratic sum of thermal motion and RMS
velocity dispersion to a reasonable accuracy.  Outliers in the plot
are due small spurious line components  sometimes introduced by the
automatic fitting program VPFIT to improve the quality of
the fit.

How realistic are the gas motions in the simulation?  To compare the
simulation to real data we have again measured a number of randomly
selected simulated absorption lines fitting Voigt profiles.  The
resulting Doppler parameter column density ($b$--log$N$) diagrams for \civ,
\ovi, and \hi\ are shown in figure \ref{bcivbovibhi}.  The scatter plots
agree very well with those from observational data
(e.g. RSWB for \civ, Hu et al.~1995 for \hi; there is no comparable
information yet for \ovi).  The presence of a minimum $b$ parameter in
both cases and a slight increase of $b$ with $N$ have both been observed.
There is also excellent  quantitative agreement with the observed
mean $b$ values: the mean \hi\ Doppler parameter in the simulations is
$\expec{b}_{\rm sim}$ = 27.3 $\kms$ ($\expec{b}_{\rm obs}$ = 28
$\kms$ is observed). The mean \civ\ Doppler parameters are 
$\expec{b}_{\rm sim}$ = 8.6 $\kms$  and  $\expec{b}_{\rm obs}$ = 9.3
$\kms$, respectively. 

Figure \ref{bsimobscomp} shows  that the shape of the simulated  
\civ\ Doppler parameter distribution  matches the observed
distribution very well. A decomposition of the measured Doppler 
parameters into thermal  and turbulent contributions gives 
a mean simulated {\it thermal} Doppler parameter  
$\expec{b}_{\rm therm} = 7.7 $\,\kms (see figure \ref{bmeasbconstr})  
with the  observed mean thermal \civ\ Doppler parameter, 
$\expec{b}_{\rm therm} = 7.2$\,\kms).

\begin{figure}[t]
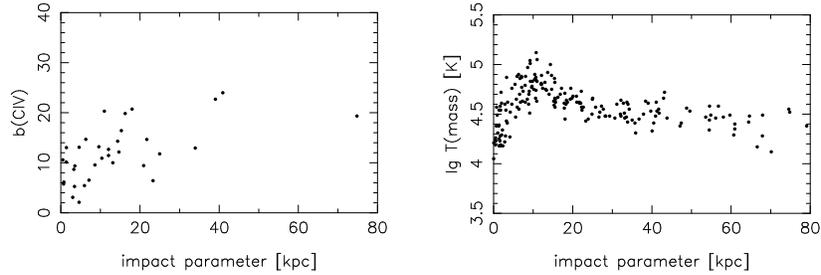

\centerline{
\psfig{file=haehnelt_17a.ps,width=5.0cm,angle=-90.}
\hspace{0.5cm}
\psfig{file=haehnelt_17b.ps,width=5.0cm,angle=-90.}
}
\vskip 0.5cm
\caption{\small Strongest \civ\ Doppler parameters of each absorption
complex as a function of impact parameter to the nearest PGC (left).
Mass-weighted temperature (right).\label{bciv_bovi_tmass_lin}}
\end{figure}

Can we observe the infalling motion during gravitational collapse
directly?  Naively we may expect to see substantial line broadening 
due to infalling gas.   However, the \hi\ density weighted 
velocity dispersion along the LOS  (the physical quantity  
relevant for the line broadening) has a median as low as 
11.6\,\kms,  even though the average radius of the regions used for 
the computation of this quantity was as large as 40 kpc.  Looking at 
the peculiar velocity diagrams in figures  
\ref{postshock} and  \ref{damped} with their large
velocity gradients of close to 200\,\kms across the size of a PGC the
quiescent structure may be somewhat surprising.  However, the enhanced
density responsible for the line formation is strongly peaked on a
spatial scale much smaller than the infalling region. The line
profile samples  mostly the high density gas which has gone through 
the shock front, has cooled to temperatures of a few $10^4$\,K,  and is more
or less at rest. In principle the signature of infalling gas is 
still visible in the form of broad profile wings. In \civ, however, 
these are almost always too weak to be seen as they are due to low 
column density gas beyond the radius of the shock.  The situation 
may be different for \hi\ absorption lines (cf. Rauch 1996). 
In \ovi\ with its much larger cross-section and weaker density 
dependence we probably also have a better chance to detect
infall. The \ovi\ Doppler parameter  ($\expec{b}_{\rm sim}$=11.5
$\kms$) is indeed higher than that of \civ, and not lower, as one
could naively expect for thermally dominated gas. Obviously
the \ovi\ absorbing gas is  subject to stronger non-thermal motions.

How do the Doppler parameters vary as a function of impact
parameter?  The LOS-integrated temperature distribution of 
either \civ\ or \ovi\ does not
show a radial variation that could be used as a measure of the impact
parameter. This is due  to the fact that some of the absorption
components are related to different PGCs. Plotting, however, only the
Doppler parameters of the  strongest absorption component in each
\civ\ complex (top panel, figure \ref{bciv_bovi_tmass_lin}) we obtain a
noticeable anti-correlation of $b$(\civ) with impact parameter from the
center of the closest PGC. This seems to reflect the drop in
temperature (bottom panel) and the coming to rest of the gas at small
radii  ($<$ 12 kpc in this case).  The trend is not obvious for \ovi\
(not shown here) which arises from a much larger region
of lower density. It also does  not show up in plots 
of $b$(\civ) when all components, rather than the strongest  
in each complex, are considered.

\begin{figure}[t]
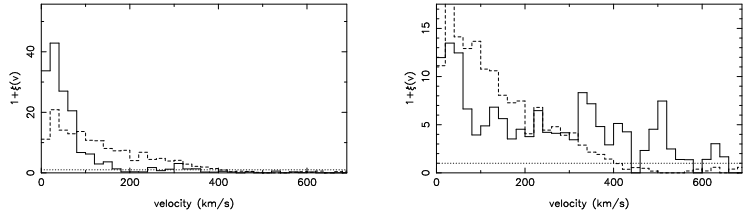

\centerline{
\psfig{file=haehnelt_18a.ps,width=4.5cm,angle=-90.}
\hspace{0.5cm}
\psfig{file=haehnelt_18b.ps,width=4.5cm,angle=-90.}
}
\vskip 0.5cm
\caption{\small Left: two-point correlation function for the \civ\ systems
detected in a simulation box at $z$=3.07 (to contain three galaxies
with  $v_c \sim 100$\,\kms at $z=0$). Normalization is by the combined TPCF 
from the observations of three  QSOs ($\expec{z}$=2.78; 
RWSB). \label{tpcf3} 
Right:  same as above for a simulation box containing a single
galaxy with  $v_c \sim 200$\, \kms\ by $z=0$.
Note the different scales of the axes.}
\end{figure}

\subsection{The two-point correlation function and  the dynamical 
state of the absorbers on large scales}

As discussed earlier the observed velocity
spreads among absorption components are difficult to reconcile with a
dynamical velocity dispersion. In paper I it was argued that  
occasional large velocity splitting of sometimes more than 
1000\,\kms\ can be explained by chance  alignments of 
the LOS with filaments containing several PGCs. The observed 
two-point correlation function
(TPCF) can indeed be interpreted in this sense, if there is a 
strong small scale clustering of \civ\ components on scales of 
20\,\kms\  in addition to the large scale expansion (RSWB).  This is in good
agreement with the earlier result that  small scale structure may
explain the ``supra-thermal'' widths commonly found for  high column 
density \op lines (Cowie et al.~1995), as well as the increasing
clustering of \op forest lines as one goes to higher
column density (Chernomordik 1995, Cristiani et al.~1995,
Fernandez-Soto et al.~1996).

The TPCF for the simulation box described earlier
(containing three $v_c \sim 100$\,\kms\
galaxies) is shown in the upper panel of figure \ref{tpcf3}. The
velocities of the \civ\ systems were obtained by Voigt profile fits to
significant continuum depressions.  
The absolute normalization comes from the observed TPCF for 
three QSOs ($\expec{z}$=2.78; RSWB).
Similar to the observed TPCF, the correlation function of 
\civ\ lines in the simulated
spectra exhibits a  narrow peak at the origin
(the lowest velocity bin is incomplete) and a long tail
out to velocities of 500-600\,\kms\ and beyond.  The  overall width  
is somewhat smaller than that of the observed distribution.
Repeating the analysis for another simulation box  containing  
one galaxy with $v_c \sim 200$\,\kms\ we
obtain the TPCF shown in the lower panel of figure \ref{tpcf3} 
which has  much more power on large scales. Considering the rather small 
size of our  simulation boxes, such variations from box to box are
not surprising.  In fact the observed TPCF is also subject to a 
large scatter from QSO to QSO, as some LOS often contain only one or 
a few systems providing large velocity splittings.
It seems likely that the observed TPCF can be explained by
averaging  over TPCFs from individual galaxy forming regions.

As a formal measure of the velocity dispersion of an individual
protogalactic region we can compute the mean (over all LOS) of the RMS of the
density weighted total velocity
\begin{eqnarray}
\sigma_v = \bigexpec{\sqrt{\frac{\int (\overline{v} -
v)^2 dn}{\int dn}}}_{\rm LOS},
\end{eqnarray}
where $v$ = $H(z)r + v_{\rm pec}$,
and
\begin{eqnarray}
\overline{v}=\frac{\int v dn}{\int dn}.
\end{eqnarray}
For the standard box, with 3 future galaxies, $\sigma_v = 100$\,\kms.
Four other boxes developing into single $\sim 200$\,\kms\
galaxies give $\sigma_v =$ 142, 134, 111 and 138\,\kms.

\subsection{Metal absorbers in emission}

The observational identification of \civ\ absorbers at redshifts $z>2$
with any known type of galaxies has proven difficult. Speculating that
the absorbers may be related to luminous galaxies with an old
population of stars Aragon-Salamanca et al.~(1994) have obtained K
images of QSO sight lines with known strong \civ\ absorbers to a K
limiting magnitude of 20.3.  A slight excess of objects probably at the
redshift  of the QSO was found but convincing identifications of
individual  absorbers were not possible.

\begin{figure}[t]
\vspace{-2.0cm}
\centerline{
\psfig{file=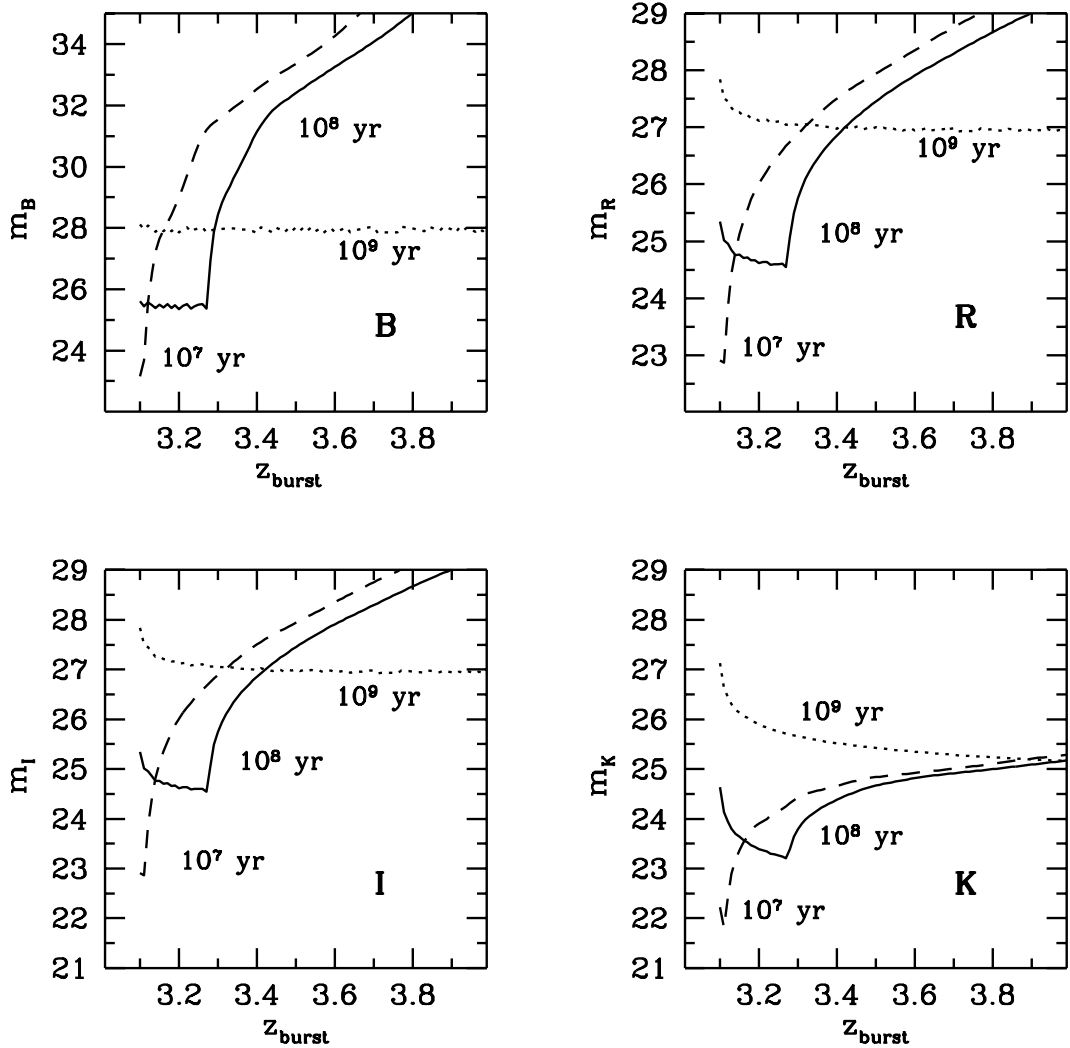,width=18.cm,angle=-0.}
}
\caption{\small
Bruzual \& Charlot models for the apparent brightness (not corrected
for \op and dust absorption, Salpeter IMF, [Z/H] = $-1.7$)
at $z=3.1$ for a total gas mass of  a protogalactic clump 
of $10^{9}$\,\Msol, as a function of the redshift where star 
formation began. Different line styles indicate  star formation 
timescales between $10^{7}$ and $10^{9}$ yr. Different panels
are for different Johnson filters. \label{martin_8}}
\end{figure}

Here we will briefly consider  stellar
population synthesis models (Bruzual \& Charlot 1993, Charlot 1996)
to investigate the prospects of detecting 
the stellar continuum from the PGCs responsible for the metal
absorption systems in our model (see also Katz 1992, Steinmetz 1996a).
We assume that the available gas turns into stars on time scales 
between $10^{7}$ and  $10^{9}$\,yr, a time span capturing the
range from a short burst to extended star formation.  In figure
\ref{martin_8} the apparent brightness (not corrected for \op and dust
absorption) at $z=3.1$ is shown for a total gas mass of  $10^{9}$\,\Msol
(typical for a PGC) and three different star formation timescales
as a function of the redshift where star formation began.

Figure \ref{martin_8} shows that our model would explain the
non-detection of \civ\ absorbers in K by  Aragon-Salamanca et al.~(1994).
Even a very short burst of $10^{9}$\,\Msol reaches only  a peak K
magnitude of 22.  Extended star formation reduces the maximum K
magnitude to 25.  A gas/star mass  $\ga 10^{11}$\,\Msol would be
necessary to be permanently detectable in K from the ground.

However, if galaxies start building up at redshifts $z\sim 3$ by
merging of smaller objects recent star formation should have occurred
at these redshifts. The optical passbands are then much more promising
for a ground-based detection of the stellar continuum.  Figure
\ref{martin_8} suggests that with  HST the stellar continuum should be
detectable in all wavebands redward of the Lyman break.  

Spectroscopic identification  will only be possible for 
the most massive  objects in the case of 
extended star formation, and for a fraction of bursting
PGCs of average mass if the star formation timescale is short. Recently
Steidel et al.~(1996) and Giavalisco et al.~(1996) have reported
the detection of a  population of star forming galaxies at comparable
redshifts.  It is intriguing that these objects could be identical to
either a bursting fraction of average mass ($10^{9}$\,\Msol) objects or
to the high mass end of the population of PGCs, showing extended star
formation (Steinmetz 1996a).

Finally one should note that confusion with foreground galaxies is a
concern when studying regions of ongoing galaxy formation as described
in this paper.  The inner 700 kpc of our simulation box (at z=3)  shown
in figures 1 and 3 correspond to an angular size slightly larger than
that of the WFPC2 field. Thus images as large as the Hubble deep field
may be required to get a ``complete'' picture of
the progenitors of an individual large $z=0$ galaxy. Superposition 
effects are then obviously quite severe.

\clearpage

\section{Discussion and conclusions}

Gravitationally driven density fluctuations  in a universe with
hierarchical structure formation can explain QSO absorption phenomena
at $z\sim 3$ over a wide range of column densities. While neutral
hydrogen shows a rather tight correlation between column density and
total density over a density range from $10^{-6}$ to $10^{-1} \cm^{-1}$,
other  ionic species probe different density and temperature
regimes in a way specific to each species. The lowest \hi\ column
densities ($10^{12}$ to $10^{14} \cm^{-2}$) arise from large-scale
sheet-like density enhancements in the IGM with an overdensity of only
a few compared  to the mean density of the universe. In this
diffuse gas (densities around $10^{-5} \cm^{-3}$) high ionization
species are prevalent and \ovi\ $\lambda  1031$ is often the
strongest metal absorption line.  Towards higher \hi\ column densities we
start probing filaments embedded in the large-scale sheets.  
In these regions low column density \civ$\lambda\lambda 1548,1550$
lines from infalling gas with densities around $10^{-4} \cm^{-2}$ 
(overdensities of about 10 to 100) dominate the metal absorption
features. \civ\ remains the most easily visible metal ion in the 
as yet unvirialized regions  around the protogalactic clumps 
which are later to merge into present-day galaxies. Still larger 
\hi\ column densities occur for lines-of-sight 
approaching the central regions of
PGCs. These give rise to Lyman limit systems and eventually to damped \op
absorbers. Total densities here exceed $10^{-4} \cm^{-3}$, and
species like \cii\ and \siiv\  become
increasingly prominent. At densities above $10^{-3} \cm^{-3}$  we have
reached the virialized region which is  generally optically thick for
radiation shortward of one Rydberg.

Although  current simulations cannot precisely constrain the size of
the damped region in a PGC, the fact that at $z \sim 3$  more than ten
such objects exist in the comoving volume  containing one $L_{\star}$
galaxy at $z=0$ considerably reduces the cross section per object 
required to explain the observed rate of incidence of damped \op absorbers.  If
this picture is correct there should be  no 1:1 correspondence between 
present-day galaxies and a high-redshift damped (or Lyman limit) absorber.
There is then no need for hypothetical large disks/halos as high
redshift progenitors of present-day galaxies.

The analysis of  artificial spectra generated from our numerical
simulation was carried out using the same methods as for observational data. The
results are in remarkable quantitative agreement with a number of
observed properties:

The predicted shape  of the \hi\ column density distribution 
shows good agreement with that of the observed distribution
and  a good fit is obtained for  
$(\Omega_b\,h_{50}^{2}/0.05)^{2}\,/(J_{-22} \, h_{50})\approx 1.5$. 
Detailed information on the strength of accompanying metal 
absorption has recently become available for \hi\ column densities 
as small as a few times $10^{14} \cm ^{-2}$. We obtain good overall agreement 
between the results from our artificial spectra and the observed properties,
either with a  simple power law UV spectrum ($\alpha$=$1.5$, $J_{-22}
=3$) or with the spectrum proposed by Haardt \& Madau (1996). This 
seems consistent with the lower end of the range of $J$ values
measured from the proximity effect  (e.g. Giallongo et al.~1996, 
and refs. therein) and suggests a baryon fraction slightly 
exceeding the nucleosynthesis constraint (Walker et al.~1991).

The scatter of the column density ratios for \siiv/\civ, \cii/\civ, and
\ovi/\civ\ versus \civ\ is consistent with the observational results,
so the  range of ionization conditions appears to be
well captured by the simulations.

As already appearent from simple photoionization models the observed
metal line strength corresponds to a {\it mean metallicity} [C/H] =
$-2.5$ for the column density range $10^{14}$ to $10^{17} \cm ^{-2}$.  
A homogeneous metal distribution reproduces the observed ion ratios 
quite well. This implies either that much of the
as yet unvirialized gas had been subject to a widespread phase of
stellar nucleosynthesis well before redshift three, or
that metal transport outward from fully  collapsed regions 
has been efficient.

Nevertheless,  the observed scatter in [C/H] is larger by a factor
three to ten than predicted  by our numerical simulations where
the metals were distributed homogeneously. This may indicate
that  some of the metal enrichment took place {\it in situ}
with incomplete  mixing prior to observation. Alternatively there
may be a wider spread in physical conditions (e.g. spatial variations of
strength and spectrum of the UV field) than assumed by the simulations.

The observed column density ratios are matched significantly better if
we use {\it relative abundances appropriate for metal-poor stars} 
(we used [C]=[N]=0 and [Si]=[O]=0.4). Damped \op systems at redshifts
$>$ 3 show  similar low metal abundances ($-2.5$ $<$ [Z/H] $<$
$-2.0$)  and relative metallicities (Lu et al.~1996, and
refs. therein).  This is consistent with the idea that  metal 
absorption systems  at high redshift  contain a record of 
early nucleosynthesis dominated by supernovae of type II.

The observed distribution of Doppler parameters and  the relative
contributions to the line width from thermal and non-thermal motion are
well reproduced by the simulations. Obviously, {\it shock heating}  is
a second important heating agent (in addition to photoionization
heating) for regions of the universe with overdensities between ten and
a few hundred.  Inspite of the large peculiar velocities ($\sim 100$\,\kms) 
of the infalling gas \civ\ absorption lines are typically only
$\sim$ 8 to 10 \kms\ wide. This is, because the \civ\ optical depth 
arises mostly in narrow post-shock regions where the shocked gas has 
already come to rest and is cooling rapidly.
The contribution of bulk motions  to the Doppler parameters 
of \ovi\ and \hi\ are larger because much of the absorption arises 
at larger impact
parameters, where infall of gas is a more important broadening agent. In
this particular model of structure formation, the \civ\ Doppler
parameter (at $z=3$) is, to a good approximation, the measure of the 
quadratic sum of the thermal  and the RMS bulk velocity dispersion.

The large scale structure in velocity space, as measured by the \civ\ TPCF,
is consistent with the observed \civ\ TPCF.  This supports the  hypothesis
proposed in paper I that LOS which intersect expanding large scale
filaments with embedded PGCs contribute significantly to the high
velocity-tail of the \civ\ TPCF.   The existence of a hypothetical
class of abundant deep potential wells at high redshift is not required.

We have discussed the prospects of detecting the stellar continuum
which is expected from protogalactic clumps if at least some 
metal enrichment has occurred in situ. These objects should be visible
at all optical wavelengths in deep images with the HST and in all optical 
passbands from the ground, at least longward of the Lyman break. 
We also suggested to interpret the Lyman break objects at
redshifts around three reported by Steidel et al.~(1996) in terms
of the high mass end of the PGCs causing metal absorption systems or of
a bursting fraction of lower mass PGCs. 
Prospects for identifying  metal absorption systems at these redshifts
are good. The best strategy may be a systematic search for 
Lyman-break objects within a few arcseconds to the line-of-sight to  
bright quasars, together with follow-up spectroscopy to the faintest 
possible limits. PGCs which are progenitors of a particular $z=0$
galaxy can be scattered over several hundred kpc at $z=3$, an area larger
than the field size of the Hubble WFPC2 camera, so problems with incompleteness and
foreground confusion can arise.

In future, a large database of metal-line ratios as a function of
redshift should improve constraints on the normalization and spectrum of
the UV background and allow us to distinguish between possible metal
enrichment histories of the IGM.  In particular,  in the
low density regions probed by the lowest column density \op absorbers,
we expect \ovi\ to be considerably stronger than \civ.  
Thus it may be possible to push metallicity determinations with \ovi\ 
closer to truly primordial gas than is possible with \civ, despite 
the severe  problems with identifying \ovi\ in the \op forest.
Another interesting case for further study is \siiv.
\siiv\ should rapidly decrease towards low densities 
and a detection in low \hi\ column density systems
would indicate that the normalization and/or  the spectrum of the UV
background differ significantly from those we have adopted.
A detection of the stellar continuum expected to be associated with
metal absorption systems  would allow us to investigate 
the radial density and  metallicity profile of the PGCs.

In summary, we have found that high-resolution hydrodynamical simulations of
galaxy forming regions can substantially aid the interpretation of the
observed properties of metal absorption systems. Currently these
simulations are best suited for studying  regions with overdensities
from ten to a few hundred. Such regions are optically thin to ionizing
radiation.  Substantial simplifications are present in the current
work. 
However, the good and sometimes excellent agreement of the present model
with observation gives reason to believe that we may have
correctly identified the  mechanism underlying many of the
metal absorption systems at high redshift; conversely, we may take 
the results presented here as an argument in favor of a hierarchical galaxy 
formation scenario.

\noindent
\section {Acknowledgments}
We thank  Bob Carswell and John Webb for VPFIT, and Gary Ferland
for making CLOUDY available to us.  Thank is also due to 
Len Cowie, Limin Lu, Andy McWilliam, Wal Sargent and Simon White 
for their helpful comments. MR is grateful to NASA for support 
through grant HF-01075.01-94A 
from the Space Telescope Science Institute, which is operated 
by the Association of Universities for Research in Astronomy, Inc.,
under NASA contract NAS5-26555. Support by NATO grant CRG 950752 and 
the ``Sonderforschungsbereich 375-95 f\"ur Astro-Teilchenphysik der 
Deutschen  Forschungsgemeinschaft'' is also acknowledged.

\pagebreak

\vfill
\break

\end{document}